\documentclass[twocolumn]{aastex6}
\usepackage{graphicx}
\usepackage[flushleft]{threeparttable}

\shorttitle{\Cii{} survey in $z>6$ quasars}
\shortauthors{Decarli et al.}

\def\Lsun{L$_\odot$}
\def\Msun{M$_\odot$}

\def\Cii{[C\,{\sc ii}]}

\def\kms{km\,s$^{-1}$}

\def\lsim{\mathrel{\rlap{\lower 3pt \hbox{$\sim$}} \raise 2.0pt \hbox{$<$}}}
\def\gsim{\mathrel{\rlap{\lower 3pt \hbox{$\sim$}} \raise 2.0pt \hbox{$>$}}}

\begin{document}

\title{
An ALMA \Cii{} survey of 27 quasars at $z>5.94$}

\author{
Roberto Decarli\altaffilmark{1,2},
Fabian Walter\altaffilmark{1,3,4},
Bram P.~Venemans\altaffilmark{1},
Eduardo Ba\~{n}ados\altaffilmark{5}, 
Frank Bertoldi\altaffilmark{7}, 
Chris Carilli\altaffilmark{4,8}, 
Xiaohui Fan\altaffilmark{9}, 
Emanuele Paolo Farina\altaffilmark{1}, 
Chiara Mazzucchelli\altaffilmark{1}, 
Dominik Riechers\altaffilmark{10}, 
Hans--Walter Rix\altaffilmark{1}, 
Michael A. Strauss\altaffilmark{11}, 
Ran Wang\altaffilmark{12}, 
Yujin Yang\altaffilmark{13}
}
\altaffiltext{1}{Max-Planck Institut f\"{u}r Astronomie, K\"{o}nigstuhl 17, D-69117, Heidelberg, Germany.}
\altaffiltext{2}{INAF -- Osservatorio Astronomico di Bologna, via Gobetti 93/3, I-40129, Bologna, Italy. E-mail: {\sf roberto.decarli@oabo.inaf.it}}
\altaffiltext{3}{Astronomy Department, California Institute of Technology, MC249-17, Pasadena, California 91125, USA}
\altaffiltext{4}{National Radio Astronomy Observatory, Pete V.\,Domenici Array Science Center, P.O.\, Box O, Socorro, NM, 87801, USA}
\altaffiltext{5}{The Observatories of the Carnegie Institution for Science, 813 Santa Barbara St., Pasadena, CA 91101, USA}
\altaffiltext{6}{Carnegie--Princeton Fellow}
\altaffiltext{7}{Argelander Institute for Astronomy, University of Bonn, Auf dem H\"{u}gel 71, 53121 Bonn, Germany}
\altaffiltext{8}{Battcock Centre for Experimental Astrophysics, Cavendish Laboratory, University of Cambridge, 19 J J Thomson Avenue, Cambridge CB3 0HE, UK}
\altaffiltext{9}{Steward Observatory, University of Arizona, 933 N. Cherry St., Tucson, AZ  85721, USA}
\altaffiltext{10}{Cornell University, 220 Space Sciences Building, Ithaca, NY 14853, USA}
\altaffiltext{11}{Department of Astrophysical Sciences, Princeton University, Princeton, New Jersey 08544, USA}
\altaffiltext{12}{Kavli Institute of Astronomy and Astrophysics at Peking University, 5 Yiheyuan Road, Haidian District, Beijing 100871, China}
\altaffiltext{13}{Korea Astronomy and Space Science Institute, Daedeokdae-ro 776, Yuseong-gu Daejeon 34055, South Korea}

\begin{abstract}
We present a survey of the \Cii{} 158\,$\mu$m line and underlying far--infrared (FIR) dust continuum emission in a sample of 27 $z\gsim6$ quasars using the Atacama Large Millimeter Array (ALMA) at $\sim1''$ resolution. The \Cii{} line was significantly detected (at $>5$-$\sigma$) in 23 sources (85\%). We find typical line luminosities of $L_{\rm [CII]}=10^{9-10}$\,L$_\odot$, and an average line width of $\sim 450$\,\kms{}. The \Cii{}--to--far-infrared luminosity ratio (\Cii/FIR) in our sources span one order of magnitude, highlighting a variety of conditions in the star--forming medium. Four quasar host galaxies are clearly resolved in their \Cii{} emission on a few kpc scales. Basic estimates of the dynamical masses of the host galaxies give masses between $2\times10^{10}$ and $2\times10^{11}$\,\Msun{}, i.e., more than an order of magnitude below what is expected from local scaling relations, given the available limits on the masses of the central black holes ($>3\times10^8$\,\Msun, assuming Eddington-limited accretion). In stacked ALMA \Cii{} spectra of individual sources in our sample, we find no evidence of a deviation from a single Gaussian profile. The quasar luminosity does not strongly correlate with either the \Cii{} luminosity or equivalent width. This survey (with typical on--source integration times of 8 min) showcases the unparalleled sensitivity of ALMA at millimeter wavelengths, and offers a unique reference sample for the study of the first massive galaxies in the universe.
\end{abstract} 

\section{Introduction} 
 
Quasars are the most luminous, non-transient sources in the early universe, and therefore represent ideal laboratories to investigate the first stages of galaxy formation. They were among the first high--redshift ($z>1$) targets for sub-mm observations \citep[e.g.,][]{omont96}. Various campaigns with bolometers mounted on single-dish telescopes \citep[e.g.,][]{bertoldi03a,beelen06,wang08a,wang08b} as well as with the {\it Herschel} observatory \citep[e.g.,][]{leipski14,drouart14} extensively sampled the far--infrared (FIR) spectral energy distributions (SEDs) of these sources. This suggested that about 1/3 of high--redshift quasars are hosted in infrared- (IR-) bright ($\sim 10^{13}$\,\Lsun{}) host galaxies \citep{leipski14}. Although the rapid accretion of material onto the central supermassive black hole may be responsible for contributing to at least part of this IR luminosity \citep[e.g.,][]{barnett15,schneider15}, it is now established that the majority of this emission is powered by star formation rates (SFRs) of several hundred solar masses per year or more in the host galaxies \citep{walter09,leipski14}. These prodigious events of star formation are fueled by immense (a few times $10^{10}$\,\Msun) reservoirs of molecular gas, which can be detected through the rotational transitions of the carbon monoxide (CO) molecule \citep[e.g.,][]{carilli02, bertoldi03b, walter03, walter04, riechers06, riechers09, wang10, venemans17b}. The gaseous reservoirs often appear spatially compact, with sizes of a few kpc or less \citep[e.g.,][]{riechers07,walter09,wang13,willott13,venemans17a}. The implied star formation rate surface densities, $\Sigma_{\rm SFR}$, can be extremely high ($\sim 1000$\,\Msun{}\,yr$^{-1}$\,kpc$^{-2}$), and might even reach the Eddington limit for star formation \citep{scoville04,thompson05,walter09}. High CO excitation is observed in some of these quasars, as would be expected in compact violent starbursts \citep[e.g.,][]{barvainis97,walter03,bertoldi03a,weiss07,riechers09,riechers11a}. 

The redshift range $z\gsim 6$ (when the age of the universe is less than 1\,Gyr) is of particular interest. The steep evolution of the average UV transmission due to neutral hydrogen's Lyman $\alpha$ line points to a rapid change in the ionization properties of the intergalactic medium, suggesting that we are entering the epoch of reionization at these redshifts \citep{fan06}. In this early phase of galaxy formation, quasar host galaxies stand out as some of the most active regions of the universe. To date (Dec 2017), there are $\sim 225$ quasars known in the literature at $z>5.5$, 119 at $z>6.0$, 15 at $z>6.5$, and only 2 at $z>7.0$ (see, e.g., \citealt{fan06}, \citealt{mortlock11}, \citealt{venemans13}, \citealt{banados16}, \citealt{jiang16}, \citealt{matsuoka16,matsuoka17}, \citealt{wang17}, \citealt{mazzucchelli17}, \citealt{banados17}, and references therein). Their black hole masses, typically estimated from the Mg{\sc ii} broad emission line at 2796\,\AA{} rest frame, are of the order of $10^9$\,\Msun{} \citep[e.g.,][]{jiang08,derosa11,derosa14}. The required rapid build-up of the black holes has sharpened our understanding of how these black holes form in the first place \citep[see][for a review]{volonteri12}. Both the gas in the broad line region around the central black hole, and the interstellar medium (ISM) of their host galaxies appear metal--enriched \citep[e.g.,][]{derosa11,mortlock11,walter03,bertoldi03a,bertoldi03b,beelen06}. This hints that the first generation of stars in these objects formed very rapidly. 

At $z\gsim6$, the fine--structure line of singly ionized carbon, \Cii{} at 158\,$\mu$m (hereafter, \Cii{}), conveniently enters the 1.2mm transparent atmospheric window, thus enabling it to be observed from the ground. The \Cii{} line is the main coolant of the cool ($<1000$\,K) ISM, in some cases reaching luminosities as high as $\sim1$ per cent of the entire far-infrared luminosity of a galaxy \citep[see, e.g.,][for a recent compilation]{diazsantos17}. Because of this, it plays a key role in the thermodynamical evolution of the ISM. Also, due to its high brightness and narrowness (as it traces the host galaxy, rather than the turbulent region close to the active nucleus), the \Cii{} line is an excellent observational tool to measure precise redshifts. These are important, for instance, in the study of the proximity zone and in the determination of the quasar lifetime (see, e.g., \citealt{fan06}, \citealt{carilli10}, \citealt{eilers17}), and to probe the gas kinematics (thus inferring the dynamical mass of the host galaxy, or exposing the presence of outflows and winds; see, e.g., \citealt{venemans12, venemans16, willott13, willott15, wang13, wang16, cicone15}). Because of its intimate connection with the cooling of the ISM, \Cii{} can also be used as a tracer of star formation \citep[e.g.,][]{delooze14,herreracamus15}. Thanks to all these virtues, \Cii{} has become a workhorse diagnostic line for the study of the ISM in galaxies at $z\gsim6$. 

In fact, the first \Cii{} line detection reported at $z>0.1$ was associated with the host galaxy of the quasar J1148+5251, at $z=6.4$ \citep{maiolino05}. This study, performed with the IRAM 30m single--dish telescope, demonstrated the feasibility of the observation of fine--structure lines (and, in particular, of \Cii{}) in $z\gsim6$ galaxies. Later studies of $z\gsim6$ quasars targeting the \Cii{} line and the underlying dust continuum capitalized on the technological improvement in sensitivity offered by the IRAM Plateau de Bure Interferometer (PdBI, now upgraded to the NOrthern Extended Millimeter Array, NOEMA; \citealt{walter09,venemans12,banados15,mazzucchelli17}) and on the advent of the Atacama Large Millimeter Array \citep[ALMA;][]{willott13,willott15,wang13,venemans16}. This decade--long effort by the community resulted in a sample of 18 quasars at $z\gsim6$ with \Cii{} detections (at various degrees of significance). 

In this paper, we mark a major step in terms of sample size. Using ALMA, we surveyed the \Cii{} line and the underlying dust continuum emission in 27 $z\gsim6$ quasar host galaxies. The size of the sample and the homogeneous data quality, together with the consistent analysis, allow us for the first time to study quasar host galaxies at the end of the reionization as a population. In addition, in \citet{decarli17} we presented a blind search for line emission in these data cubes that resulted in the detection of \Cii{}--bright galaxy companions in four of the quasars in our sample. These companion galaxies offer a first insight on the properties of the close galactic environment of $z\gsim6$ quasars. The focus of the present paper is on the line properties of the quasar host galaxies themselves, while a companion paper (Venemans et al.~in prep.) will analyze the underlying continuum emission from the same dataset. 

The paper is organized as follows: In Section \ref{sec_survey} we outline our survey and the reference sample from the literature; in Section \ref{sec_observations} we describe the observations and the data reduction; in Sections \ref{sec_results} and \ref{sec_analysis} we present the measurements and infer derived quantities, respectively. Finally, we discuss our results and draw our conclusions in Section~\ref{sec_discuss}.

Throughout the paper we assume a standard $\Lambda$CDM cosmology with $H_0=70$ km s$^{-1}$ Mpc$^{-1}$, $\Omega_{\rm m}=0.3$ and $\Omega_{\Lambda}=0.7$ \citep[consistent with the measurements by the][]{planck15}. Magnitudes are reported in the AB photometric system.

\section{The \Cii{} survey}\label{sec_survey}

\subsection{The main sample}

\begin{figure}[h]
\begin{center}
\includegraphics[width=0.99\columnwidth]{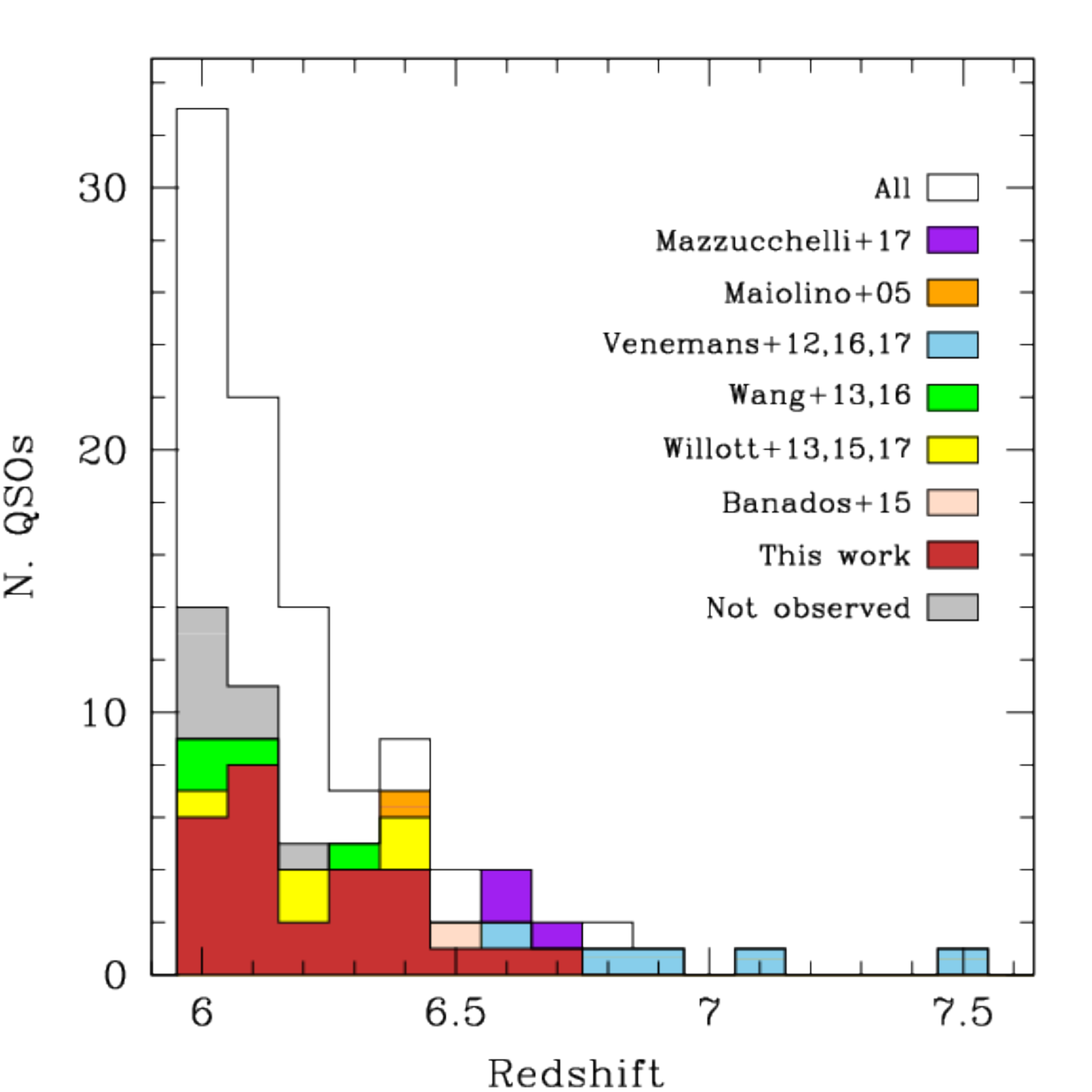}\\
\end{center}
\caption{Redshift distribution of the quasars in our sample (observed: dark red; unobserved: grey), compared with the other $z\gsim6$ quasars with \Cii{} observations (filled histograms -- see the legend for references) and the parent sample of all known $z\gsim6$ quasars. }
\label{fig_z_distr}
\end{figure}

The parent sample of our ALMA survey was designed to include all the known quasars matching the following criteria: 
\begin{itemize}
\item[1)] they lie at $z>5.94$ (i.e., $\nu_{\rm obs}$(\Cii) $< 273.854$\,GHz), so that the \Cii{} line falls in ALMA band 6 ($\sim 1.2$mm);
\item[2)] they are at declination Decl.~$<+15^\circ$, so that they can be observed at high elevation from the ALMA site;
\item[3)] they are more luminous than $M_{1450}$=$-25.25$\,mag, where $M_{1450}$ is the absolute magnitude derived from broad-band imaging observations of the rest-frame FUV continuum.
\item[4)] they were not previously targeted in the \Cii{} line.
\end{itemize}
At the time of the ALMA Cycle 3 deadline (April 2015), there were 57 published quasars at $z>5.94$, 39 of which match the declination requirement. Two of these objects do not match the FUV luminosity requirement\footnote{The object J1152+0055 falls just below our $M_{\rm UV}$ cut based on the Pan-STARRS y-band, but it is above the cut using the J-band photometry. We therefore included this object in the sample.}. Eleven of the remaining 37 sources had been observed in \Cii{} in past programs, either with ALMA or other facilities (mostly, IRAM/PdBI). We also include 9 new quasars that match our selection criteria but were still unpublished at the time the proposal was submitted \citep[][Venemans et al.~in prep.]{banados16,mazzucchelli17}. The final sample therefore consists of 35 sources (see Table~\ref{tab_sample}). Observations were performed for 27 of them, chosen only due to their visibility at the time of the observations. 
For reference, Figure~\ref{fig_z_distr} shows the redshift distribution of the quasars in our sample, compared to all the other quasars currently known in this redshift range.

\subsection{The sample from the literature}

Our results are complemented with all the \Cii{} and underlying dust continuum observations of $z>5.94$ quasars available in the literature. This sample consists of 1 quasar from \citet{maiolino05} and \citet{walter09}; 1 quasar from \citet{banados15}; 5 quasars from \citet{venemans12,venemans16,venemans17a,venemans17c}; 5 from \citet{willott13,willott15,willott17}\footnote{One of the quasars discussed in \citet{willott17}, PSO J167--13, was independently observed as part of our survey; we hereby refer only to our observations for the sake of a homogeneous comparison. The two datasets produce consistent results.}; 4 quasars from \citet{wang13,wang16}; and 3 from \citet{mazzucchelli17}. The sample from the literature is presented in Table~\ref{tab_lit}. For these sources, we will use the line luminosity and width, and the dust continuum flux density derived in the papers quoted above. For consistency with the rest of our sample, however, we re-derive the IR luminosity using the same approach adopted for the main sample of this work. 

\begin{figure*}
\includegraphics[width=0.99\textwidth]{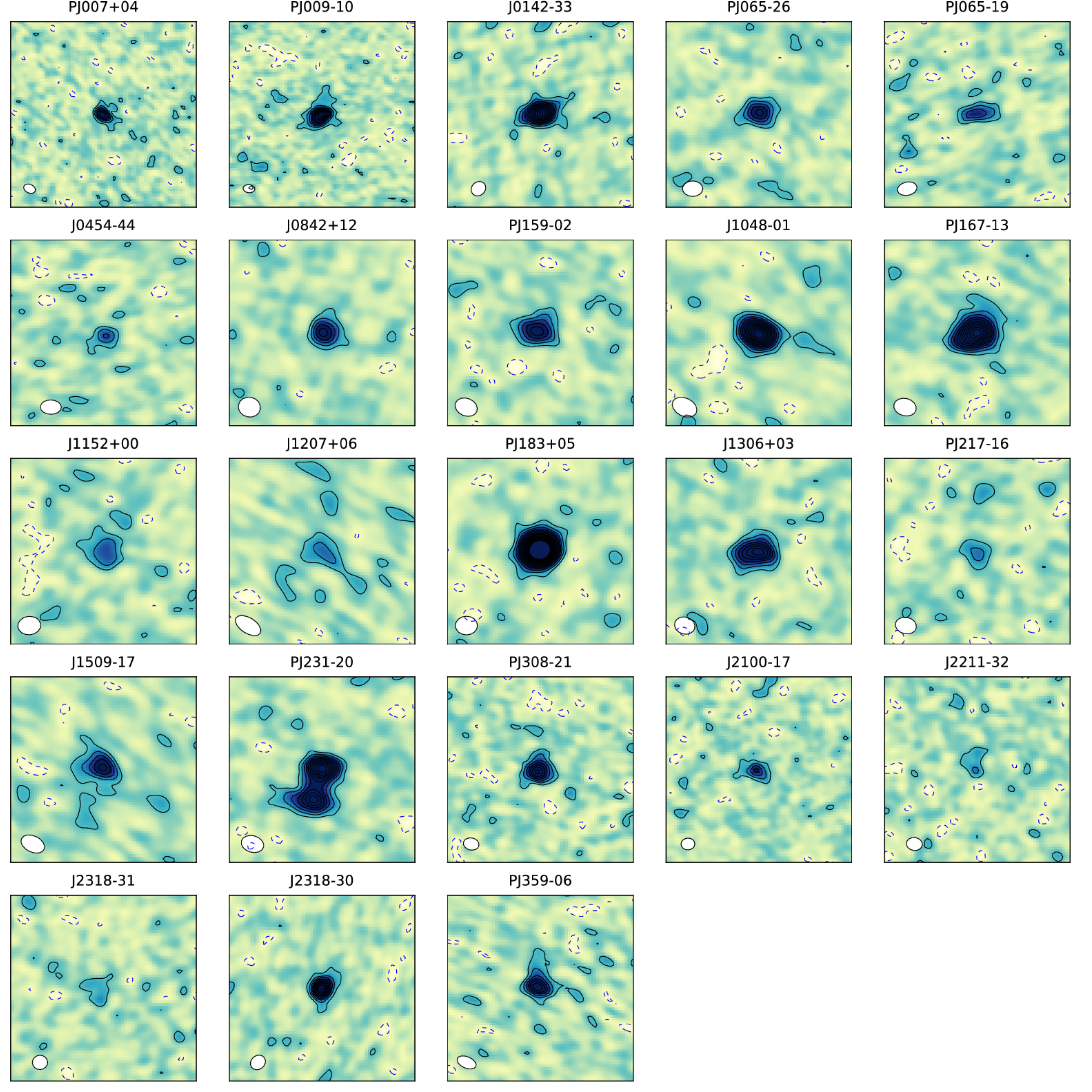}\\
\caption{ALMA postage stamps showing the continuum-subtracted \Cii{} line maps of each of the 27 sources in our sample, integrated over a width of $\pm1.4\times\sigma_{\rm line}$, in order to maximize the line S/N. Each panel is $10''\times10''$ wide. North is up, East to the left. Only quasars with a \Cii{} detection are shown. The solid black / dashed blue contours mark the $\pm 2,4,6,$\ldots $\sigma$ isophotes. The synthesized beam of the observations is shown in the bottom-left corner of each panel. Extended names are reported in Table~\ref{tab_sample}.}
\label{fig_alma_ps}
\end{figure*}

\begin{figure*}
\begin{center}
\includegraphics[width=0.89\textwidth]{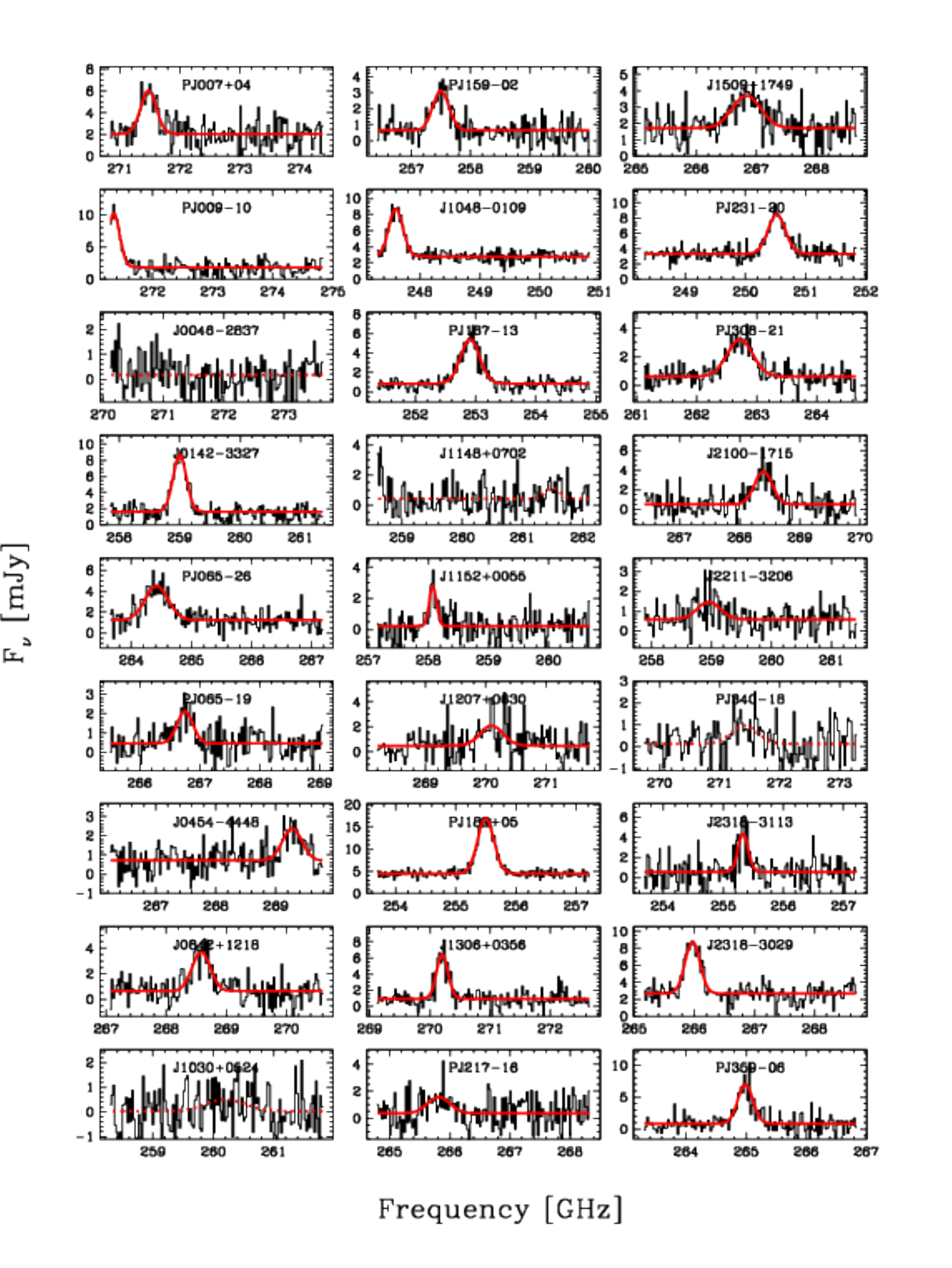}\\
\end{center}
\caption{ALMA spectra of \Cii{} and underlying continuum of the quasars in our sample (black histograms), plotted with a 30\,\kms{} velocity binning. The best fit Gaussian line+flat continuum models are shown as solid red lines. The fit values for continuum and line are summarized in Table~\ref{tab_spc}. We also show with dotted lines the best fits in those cases where no significant line emission is detected.}
\label{fig_alma_spc}
\end{figure*}

\section{Observations, data reduction and analysis}\label{sec_observations}

The ALMA data set (program ID: 2015.1.01115.S) used in this study consists of short ($\sim8$\,min on source) pointings centered on the optical/NIR coordinates of the quasars. The tuning frequency of the Spectral Windows (SPWs) was chosen as follows: Two SPWs (0 and 1) encompassed the expected observed frequency of \Cii{}, given our best pre--ALMA redshift estimate (see Table~\ref{tab_sample}), with a small overlap (typically $<5$\% of the bandwidth) between SPW 0 and 1 to account for noisy edge channels. This results in a frequency coverage of $\sim 3.6$\,GHz around the expected observed frequency of the \Cii{} line, or $\Delta z\approx 0.014$. At $z=6.5$, the \Cii{} line ($\nu_0=1900.548$\,GHz) is shifted to $\nu_{\rm obs}=253.406$\,GHz. The other two SPWs (2 and 3) covered rest-frame frequencies around 1790\,GHz. Observations were carried out between 2016 January and July, with the array in compact configuration with 38-49 12\,m antennas. The primary beam of the 12m ALMA antennas is $\sim25''$ in diameter at 250\,GHz. The synthesized beam size is typically $\sim 1.1''\times 0.8''$. The typical rms noise is 0.5\,mJy\,beam$^{-1}$ per 30\,\kms{} channel. Table~\ref{tab_obs} provides details on the observations, beam size, and sensitivity reached for each target.

We processed the data using the CASA \citep{mcmullin07} pipeline for ALMA (version 4.7.1), using the default calibration procedure. The cubes were imaged using Briggs cleaning (via the CASA task \textsf{tclean}) with robustness parameter = 2 (i.e., natural visibility weights) to maximize the signal-to-noise ratio of our observations. The imaging process involves the following steps: 
\begin{itemize}
\item[1)] First, we collapse SPWs 0 and 1 into a single continuum+line map for each target. This map is inspected in order to identify the position of detected sources, and to estimate the depth of the observations (via a 3-$\sigma$ clipped estimate of the map rms). A box mask is created around each detected source. 
\item[2)] We then image the data cube (coupling SPWs 0\&1 and 2\&3), using the mask to identify the cleaning regions, and adopting a 2-$\sigma$ cleaning threshold based on the continuum estimate (assuming roughly constant noise per channel throughout the cube). We adopt 30\,\kms{} channels with a linear interpolator in order to resample the cube in velocity space.
\item[3)] We extract 1D spectra on a single-pixel basis centering at the position of the detected sources. 
\item[4)] We fit the extracted spectra with a flat continuum + a Gaussian. 
\item[5)] In order to create the line map, we collapse the data cube in the frequency range set by the line peak $\pm 1.4\,\sigma_{\rm line}$ from the best (spectral) fit. Here, $\sigma_{\rm line}$ is the line width from the Gaussian fit, giving a Full Width at Half Maximum FWHM$\approx$2.35 $\sigma_{\rm line}$. This choice maximizes the S/N of the line map, in the case of a perfectly Gaussian line profile, and constant noise. For a Gaussian line profile, integrating within $\pm1.4\,\sigma_{\rm line}$ recovers 83\% of the total line flux. 
\item[6)] We create a pure--continuum map by using the line--free channels in SPW 0\&1, or the entire available bandwidth for SPWs 2\&3. We opt not to combine all the continuum maps, in order to preserve information about the spectral slope.
\item[7)] The pure--line map is obtained by subtracting (in uv-space) the continuum emission from the map created at step \#5. 
\end{itemize}

The continuum-subtracted line map of each target in our study is shown in Figure~\ref{fig_alma_ps}.

\section{Results}\label{sec_results}

For sources that are unresolved, or only marginally resolved, two different approaches can be used in order to infer a line flux. The first one is to extract the spectrum at a single spatial pixel (from step 3 of the imaging procedure described above), and fit the line profile along the frequency / velocity axis. The second approach consists of creating a line map, and fit the emission in the sky plane (as described in steps 5-7 of the imagine procedure). The first approach has the advantage that the full spectral information is taken into account (including potential wings in the line profile); on the other hand, it might suffer from flux losses if the emission is spatially extended. In contrast, the second approach captures the spatial extent of the line emission, but misses the wings along the spectral (i.e., velocity) dimension of the cubes. Here we adopt both approaches, and then estimate the impact of the underlying assumptions in order to establish the best estimate of the \Cii{} line emission in the sources in our sample.

\subsection{\Cii{} spectral measurements}\label{sec_spc}

\begin{figure}
\begin{center}
\includegraphics[width=0.99\columnwidth]{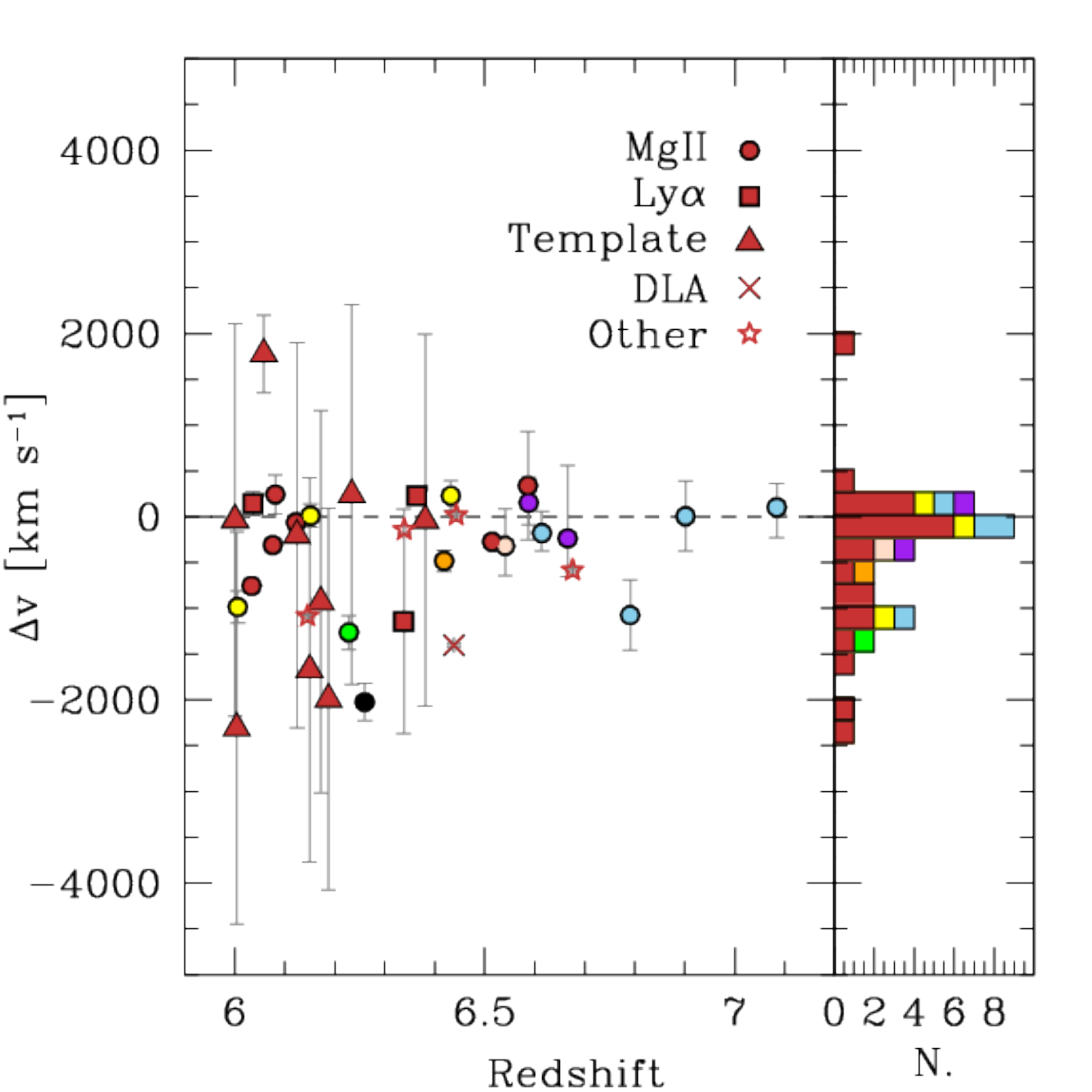}\\
\end{center}
\caption{Velocity offset between the pre-ALMA redshift estimates of the quasars, and their \Cii{}--based redshifts ($v_{\rm pre-ALMA}-v_{\rm [CII]}$). The different methods used in the pre-ALMA redshift estimates (mostly based on features associated with the broad--line region) are indicated by different symbols, while the color-coding highlights the reference for the \Cii{} redshift using the same color--scheme of Figure~\ref{fig_z_distr}. While part of the scatter is due to the large uncertainties in the pre-ALMA methods, we find significant shifts of several hundred \kms{} in several quasars. }
\label{fig_dv}
\end{figure}

\begin{figure}
\begin{center}
\includegraphics[width=0.99\columnwidth]{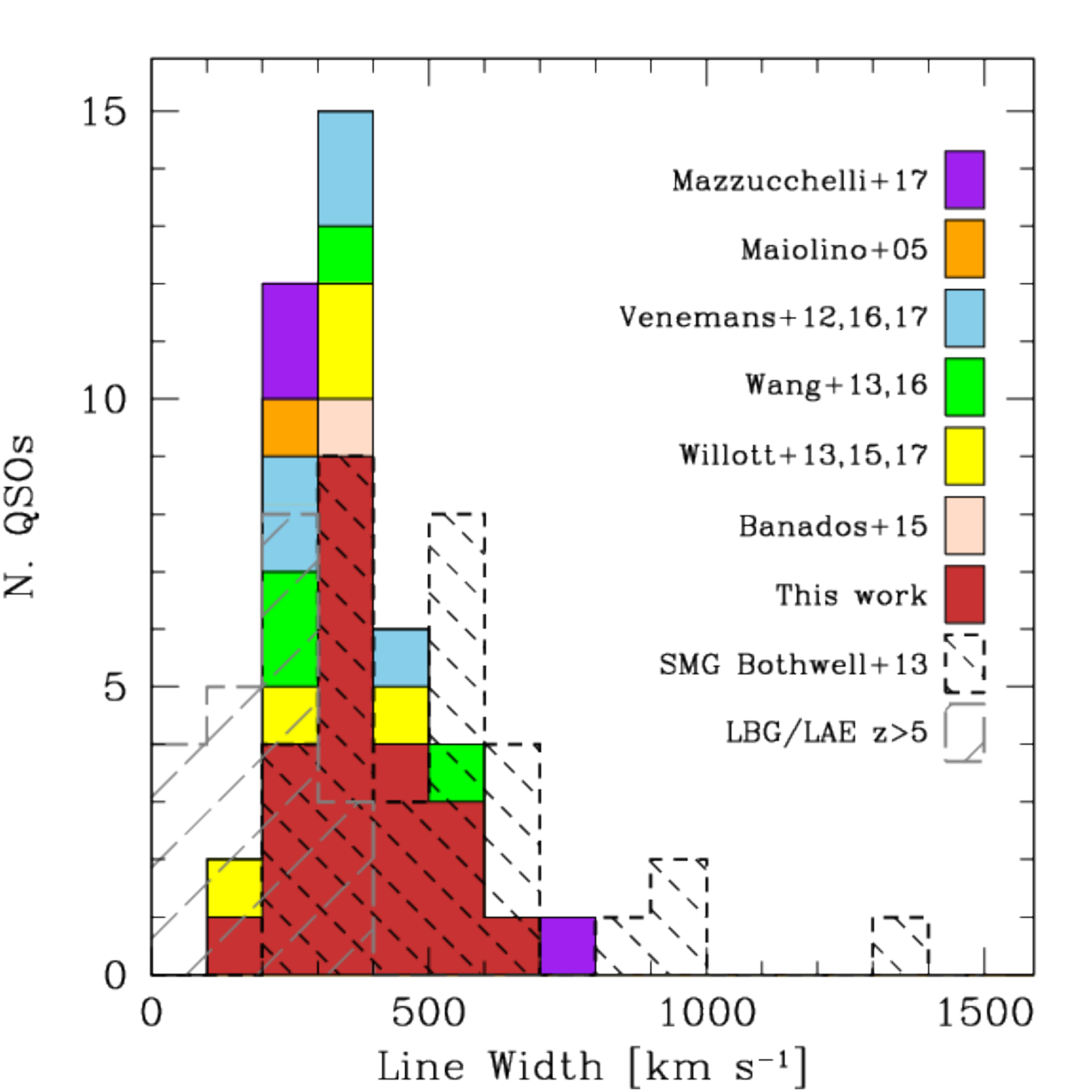}\\
\end{center}
\caption{Distribution of the line width (FWHM) of \Cii{} in the quasars of our sample, compared with the literature sample (indicated with different colors), and with the distribution of CO line widths in sub-mm galaxies from \citet{bothwell13} and of \Cii{} line widths in $z>5$ Lyman Break Galaxies and Ly$\alpha$ Emitters \citep{riechers14,capak15,maiolino15,pentericci16}. The distributions of quasars and SMGs are statistically indistinguishable, whereas LBGs/LAEs show narrower lines.}
\label{fig_width}
\end{figure}

\begin{figure*}
\begin{center}
\includegraphics[width=0.99\columnwidth]{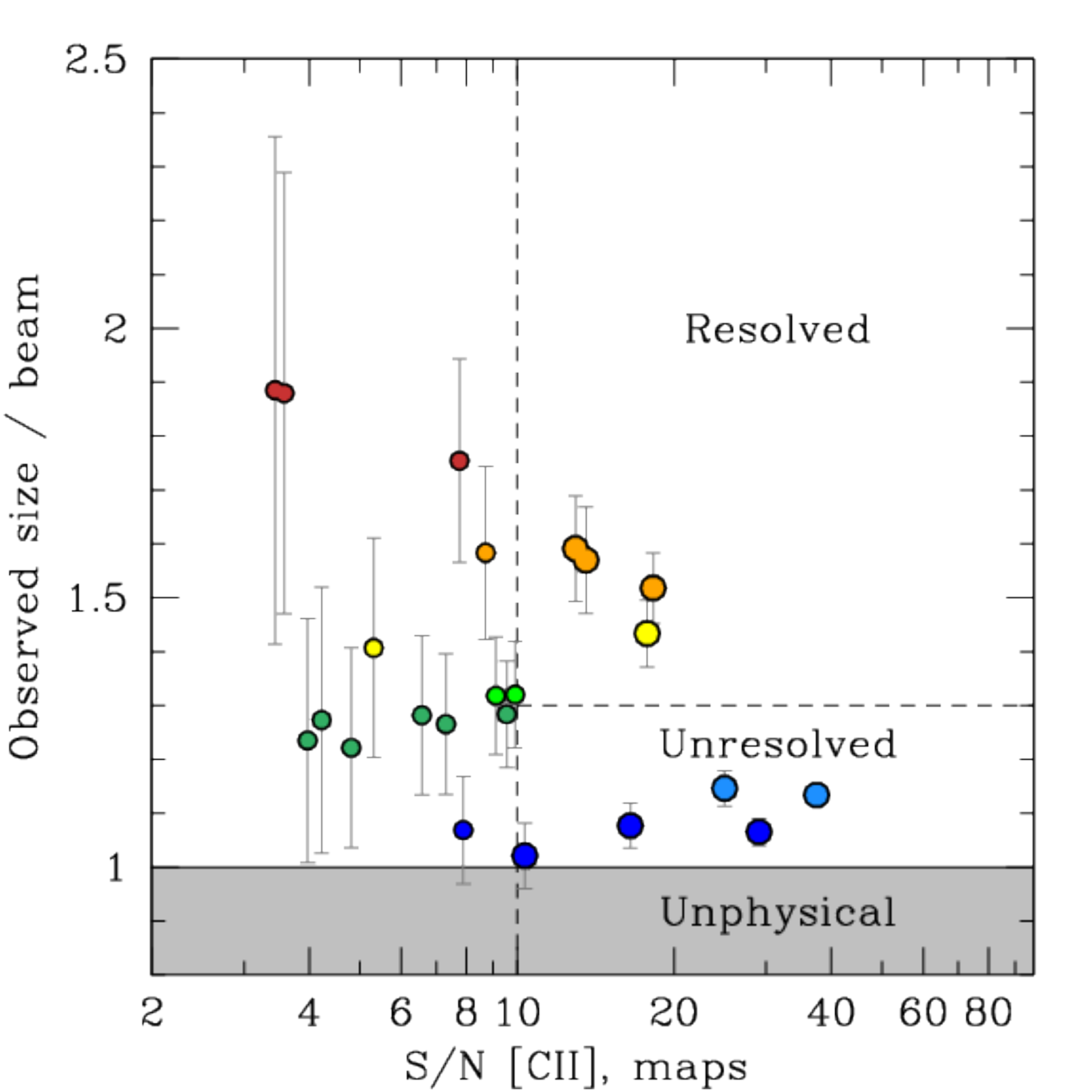}
\includegraphics[width=0.99\columnwidth]{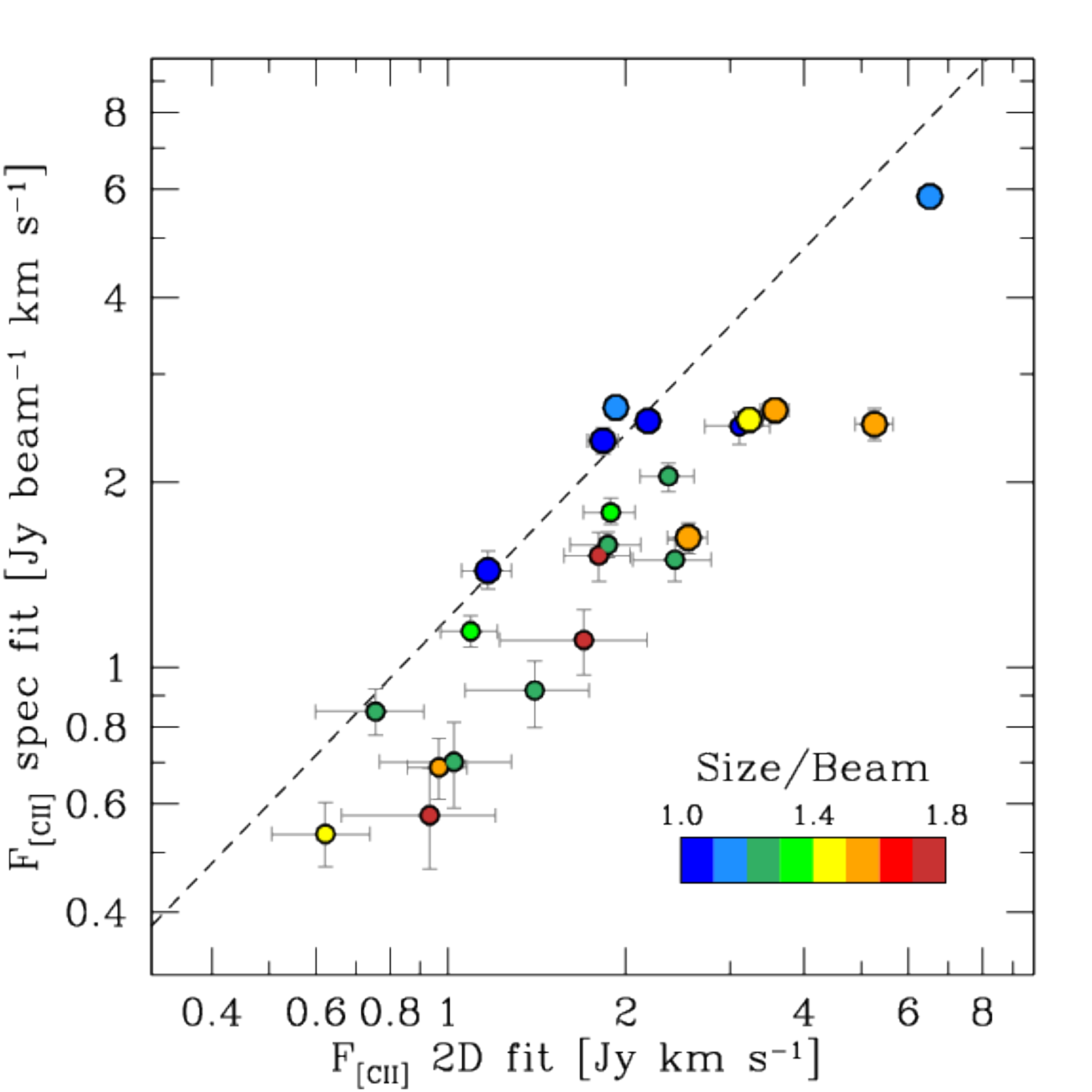}\\
\end{center}
\caption{Size estimate of the \Cii{}--emitting region and impact on single--pixel extractions. {\em Left:} The observed (i.e., beam-convolved) size estimated from the 2D Gaussian fit of the continuum--subtracted \Cii{} line maps, plotted as a function of the signal--to--noise ratio of the line detection, as estimated from the maps. The color code of the points reflects the y-axis position of the points, while the bigger symbols highlight the objects with S/N$>$10. {\em Right:} The \Cii{} flux measured from the single--pixel extraction of the spectra, and fitted with a Gaussian line profile (see Figure~\ref{fig_alma_spc}), as a function of the \Cii{} flux measured with the 2D Gaussian fit of the line maps. The dashed line shows the one-to-one case, after applying a correction for the wings of the lines that have not been included in the maps. The symbol size and color scheme are the same as in the left--hand panel.}
\label{fig_beam_ima_spc}
\end{figure*}
As described in the previous section, we first searched the collapsed continuum+line maps for sources, and extrated spectra at the position corresponding to the emission peak. This approach is justified as the size of the \Cii{}--emitting regions in quasar host galaxies is comparable with the resolution element of our sources \citep[][see also Figure~\ref{fig_beam_ima_spc} in the present work]{venemans17a}. In those objects in which no clear detection associated with the quasar host galaxy was found, we extracted the spectra at the nominal position of the target based on optical/NIR data. The spectra, extracted over SPWs 0\&1, of all the quasars in our study are shown in Figure~\ref{fig_alma_spc}.

We fit the spectra assuming a flat continuum emission from the dust, plus a Gaussian for the \Cii{} line. We use a custom Metropolis Monte Carlo Markov Chain code to sample the posterior probability of the models, given the observed data. We adopt a flat prior for the line peak frequency, a Maxwellian distribution with a line width of 300\,\kms{}, and a broad Gaussian distribution as prior for the flux density of both the line peak and the continuum. The fits and their uncertainties at 1-$\sigma$ significance are derived as the median and the 14\%--86\% quartiles of the posterior distribution. The resulting line parameters (peak frequency, width, flux) are reported in Table~\ref{tab_spc}. We consider a line detected if its integrated flux exceeds 5 times its lower-side uncertainty. Out of 27 targeted quasars, 4 do not match this criterion: J0046-2837, J1030+0524, J1148+0702, and PJ340-18. The depths of these observations were not significantly different from the remainder of the sample. 

Figure~\ref{fig_dv} shows the velocity difference between the pre--ALMA redshift estimates and their \Cii{}--based redshifts. There is significant scatter, largely due to the rather large uncertainties of some pre--ALMA redshift estimates. This is particularly true for redshift estimates based on the Ly$\alpha$ features (in emission and/or absorption) only, or for quasars for which we only have optical spectroscopy (thus sampling only a few, generally fainter lines). A number of sources, however, show statistically--significant velocity differences of several hundred \kms{} (and up to a few thousand \kms), with a tendency towards blue-shifts of the pre--ALMA estimators. This supports earlier evidence for such shifts \citep[e.g.,][]{riechers11b,venemans16}, but on a larger, homogeneous sample. The weighted average of the velocity offsets (including both our sample and the reference literature sources) is $-620\pm8$\,\kms{} (accounting for the formal uncertainties in both the redshift estimates). Given that the pre--ALMA estimates are mostly based on features of the broad-line region, it is possible that these shifts are due to outflowing material or winds close to the central black holes.

The large differences observed between pre--ALMA and \Cii{} redshifts may be responsible for at least part of the \Cii{} non detections in our survey. The combination of SPWs 0\&1 in the ALMA data results in a coverage of $\sim 4100$\,\kms{} at $z=6.0$. In fact, in the case of PJ009--10, J0454--4448, and J1048--0109, the \Cii{} line is observed at the edge of the band, so it is possible that in some cases the line has just been missed by our observations. This is likely the case for J1148+0702, the redshift of which was revised after our observations were performed \citep{jiang16}. The other non--detections (J0046--2837, J1030+0524, and PJ340--18) also show the faintest dust continua in our survey, thus suggesting that they might be intrinsically faint in \Cii{}.

Figure~\ref{fig_width} compares the distribution of \Cii{} line widths (expressed as FWHM of the Gaussian fit) for the quasars in our sample, to the literature sample. The mean and standard deviations of the two distributions are $385\pm115$\,\kms{} and $350\pm125$\,\kms{}, respectively. A Kolmogorov-Smirnov test suggests that the parent distribution from which the two samples are drawn is statistically indistiguishable. This is not surprising, as the majority of the sample of quasars from the literature fulfill both the redshift and the UV luminosity criteria used in the definition of our sample (see Section~\ref{sec_survey}). Similarly, we compare the width distribution of \Cii{} in $z>6$ quasars with the one of CO in sub-mm galaxies at $z=1-3$ from \citet{bothwell13}. The latter shows a tail towards broader line widths compared with the former, but the difference is not statistically significant. On the other hand, Lyman Break Galaxies (LBGs) and Ly-$\alpha$ Emitters at $z>5$ \citep{riechers14,capak15,maiolino15,pentericci16} show significantly narrow \Cii{} lines, with a mean and standard deviation of $200\pm100$\,\kms{}.


\subsection{\Cii{} 2D measurements}\label{sec_cii2d}

We can also infer the line flux and the size of the \Cii{}-emitting region via a fit of the continuum-subtracted maps shown in Figure~\ref{fig_alma_ps}. The fit is performed within CASA, selecting a narrow rectangular region around the quasars for the fit. Because the maps have been created by integrating over $2.8\times\sigma_{\rm line}$, for a gaussian line profile, only $\sim 83$\% of the total line emission is enclosed in the maps -- this factor is taken into account in our flux measurements. The intrinsic emission is modeled as a 2D-gaussian profile, with the centroid position, integrated flux, deconvolved major and minor axis, and the position angle as free parameters. The modeled emission is then convolved with the observed beam, when fitting to the data. The derived parameters are listed in Table~\ref{tab_size}. 

The 2D Gaussian fits suggest that the sizes of the \Cii{} emission are comparable to (and at most, twice as large as) the synthesized beams, suggesting that our \Cii{} size estimates are only tentative. In Figure~\ref{fig_beam_ima_spc} we assess the impact of extended emission in our \Cii{} flux estimates. First, we plot the ratio between the observed (i.e., beam--convolved) size and the beam major axis, as a function of the signal--to--noise ratio (S/N) of the \Cii{} line as measured from the 2D fit. At S/N$>$10, we find 5 sources with compact \Cii{} emission (observed size $\approx$ beam), and 4 with resolved emission on scales of $\sim 1.5\times$ the beam size. This difference in size is also reflected in the comparison between the \Cii{} flux estimated from the single-pixel extraction of the spectrum, and the one based on the 2D Gaussian fit (see Figure~\ref{fig_beam_ima_spc}, {\em right}): the more extended objects tend to deviate from the 1--to--1 relation. In the following, we define a source's \Cii{} flux through the measurement via the 2D spatial Gaussian fit, and scaling by a fixed $1/0.83\times$ factor to account for the flux associated with the wings of the line that are not included in the maps.

In \citet{decarli17}, we presented the search for companion \Cii{}--bright sources in the field of our observations. Out of 25 fields with comparable depth (i.e., excluding J2318-31 and J2318-30 because of inferior data quality), we found companion \Cii{}--bright galaxies in 4 cases (J0842+1218, J2100--1715, PJ231--20, and PJ308--21). In two cases, the separation between the companion and the quasar host galaxy is $\sim 10$\,kpc, or $\sim 2''$. Given the angular resolution of our data, it is possible that more companion galaxies are present at very small angular separation ($\sim 1''$) from other quasars \citep[in particular, see the discussion on P167--13 in][]{willott17}. This would clearly affect our estimates of the size, surface brightness, and integrated luminosity of these sources.

\section{Analysis}\label{sec_analysis}

\begin{figure}
\begin{center}
\includegraphics[width=0.99\columnwidth]{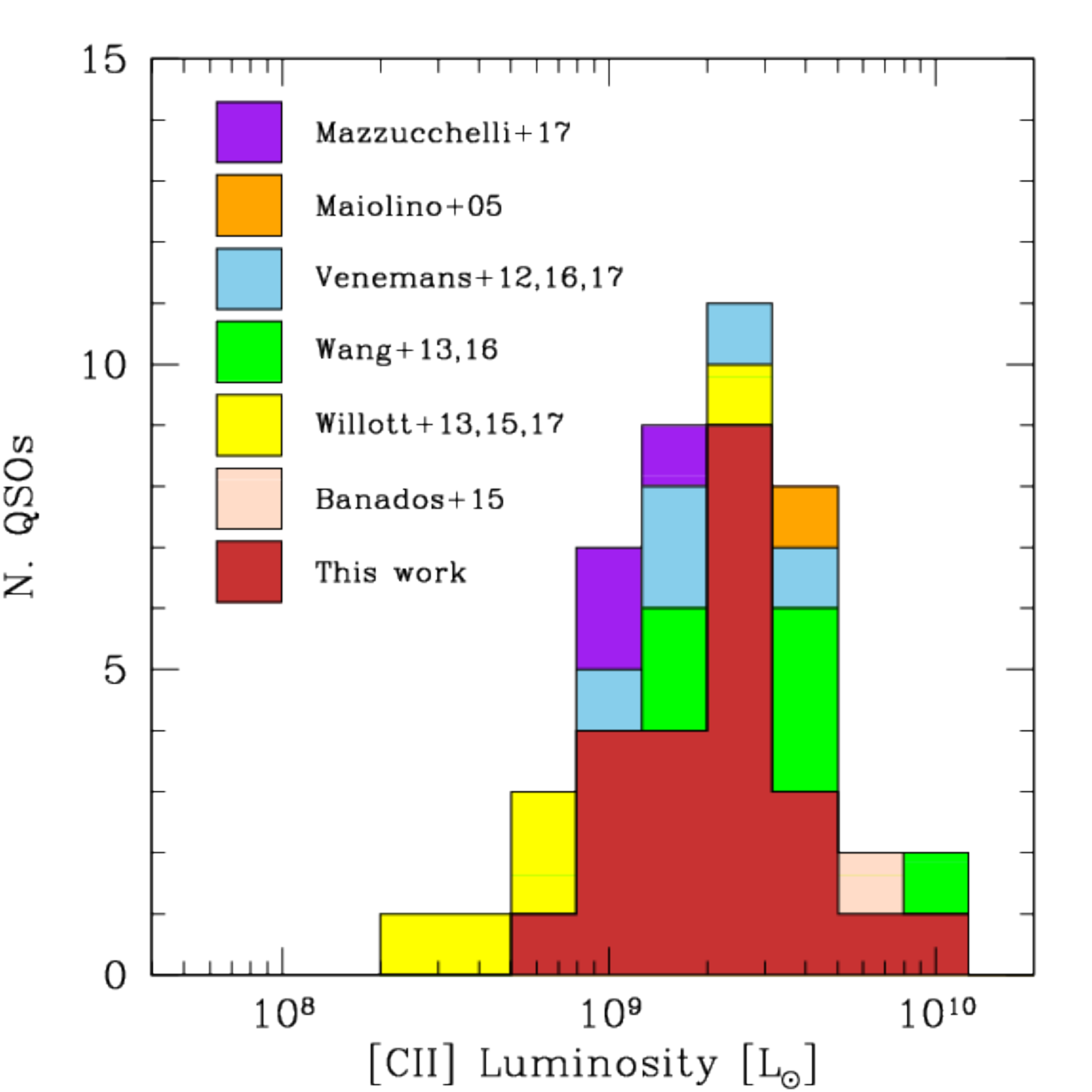}\\
\end{center}
\caption{Distribution of \Cii{} luminosity in our sample and in the sample from the literature. Typical line luminosities are in the range $(1-5)\times10^9$\,\Lsun{}. Only sources detected at $>$5-$\sigma$ (in the spectral fit, see Section~\ref{sec_spc}) are shown.}
\label{fig_lum_distr}
\end{figure}

\subsection{\Cii{} and continuum luminosities}

The line fluxes are converted into line luminosities following:
\begin{equation}\label{eq_line_lum}
\frac{L_{\rm [CII]}}{\rm L_\odot}=1.04\times10^{-3}\, \frac{F_{\rm line}}{\rm Jy\,km\,s^{-1}}\,\frac{\nu_{\rm obs}}{\rm GHz}\, \left(\frac{D_{\rm L}}{\rm Mpc}\right)^2,
\end{equation}
where $F_{\rm line}$ is the integrated line flux (from the 2D fit of the line described in Section \ref{sec_cii2d}, corrected for the flux loss due to the line wings), $\nu_{\rm obs}$ is the observed frequency of the line, and $D_{\rm L}$ is the luminosity distance \citep[see, e.g.,][]{carilli13}. Figure~\ref{fig_lum_distr} shows the distribution of \Cii{} luminosity for the quasars in our main sample, and in the literature sample. The \Cii{} lines of the quasars in our sample span over a dex in luminosity, with a peak in the distribution around $3\times 10^9$\,\Lsun{}. This is consistent with the range of luminosities typically reported in similar sources from previous studies (see also Table~\ref{tab_lit}), with the exception of the \citet{willott13,willott15} sample which focuses on sources fainter in the rest-frame UV and in the \Cii{} emission than the ones selected here.

\begin{figure*}
\begin{center}
\includegraphics[width=0.99\columnwidth]{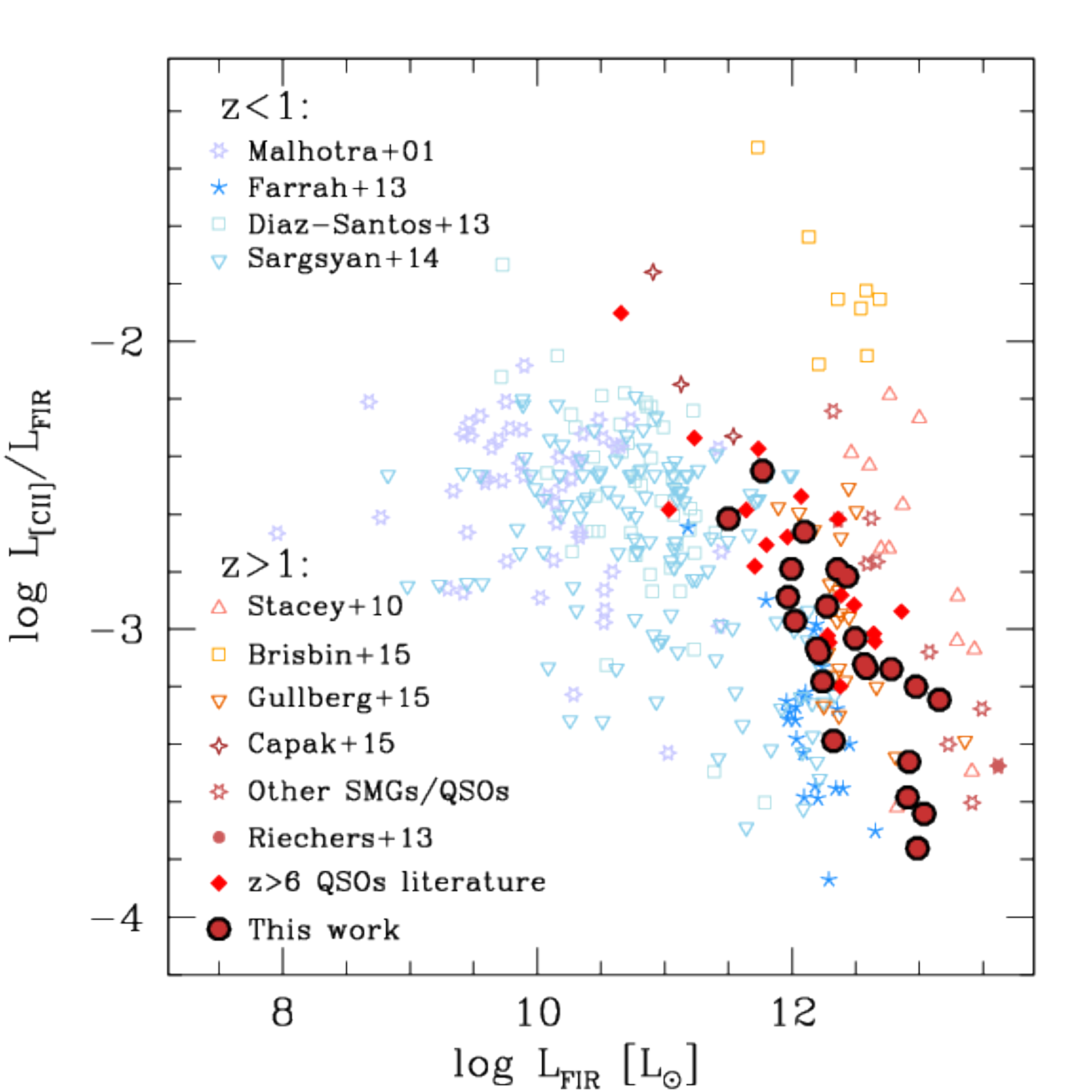}
\includegraphics[width=0.99\columnwidth]{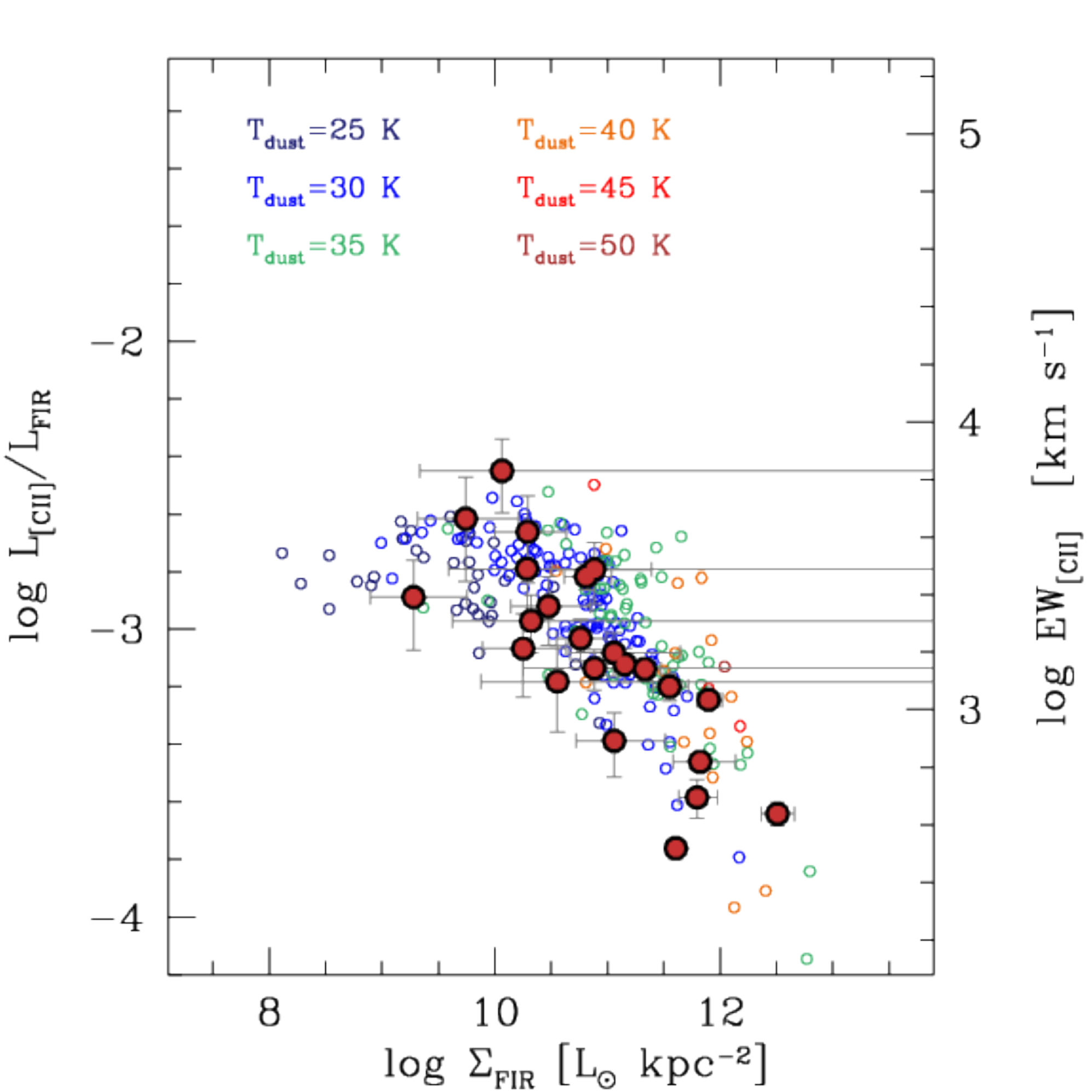}\\
\end{center}
\caption{{\em Left:} The \Cii{}--to--FIR luminosity ratio as a function of FIR luminosity, for the sources in our sample as well as a compilation of $z<1$ sources (blue/cyan symbols; data from: \citealt{malhotra01}, \citealt{farrah13}, \citealt{diazsantos13}, \citealt{sargsyan14}) as well as high-$z$ objects (red/orange symbols; data from: \citealt{stacey10}, \citealt{brisbin15}, \citealt{gullberg15}, \citealt{capak15}, \citealt{riechers13}, \citealt{riechers14}, \citealt{sargsyan14}). For all the samples, we do not plot upper limits for the sake of clarity. Quasars in our sample span a wide range (over a dex) of \Cii{}/FIR ratio, ranging from ULIRG--like values (\Cii{}/FIR$\sim$0.001) to values closer to local star-forming galaxies (\Cii{}/FIR$\sim$0.01). This highlights the diversity in the global properties of the ISM in $z>6$ quasars. {\em Right:} The \Cii{}/FIR ratio as a function of the FIR luminosity surface density, $\Sigma_{\rm FIR}$. Empty circles show the GOALS IR--luminous galaxies from \citet{diazsantos17}, color--coded based on the dust temperature estimated from the $F_\nu$(63\,$\mu$m)/$F_\nu$(158\,$\mu$m) ratio. }
\label{fig_cii_fir}
\end{figure*}

By comparing the \Cii{} emission with the underlying continuum, we gain important insights into the physical properties of the ISM in our sources. The most commonly used diagnostic in this context is the ratio between the \Cii{} luminosity and the integrated luminosity of the underlying dust continuum. A detailed discussion of the dust continuum properties of the quasars in our sample is deferred to a companion paper (Venemans et al.~in prep). Briefly, in order to infer IR luminosities, we model the dust continuum emission as a modified black body \citep[see, e.g.,][]{dunne00,beelen06}:
\begin{equation}\label{eq_mbb}
L_\nu({\rm dust}) = \frac{2 h \nu^3}{c^2}\, \kappa_\nu(\beta)\,\frac{M_{\rm dust}}{e^{h \nu/k_{\rm b} T_{\rm dust}}-1}
\end{equation}
where $T_{\rm dust}=47$\,K is the dust temperature, $\kappa_\nu(\beta)=0.77\,(\nu / {\rm 352\,GHz})^\beta$\,cm$^2$\,g$^{-1}$ is the opacity law, and $\beta=1.6$ is the (dust) emissivity index. {The values of $T_{\rm dust}$ and $\beta$ assumed here are taken from \citet{beelen06}, and are consistent with similar, more recent studies \citep[e.g.,][]{leipski14}\footnote{The use of a single modified black body component might lead to a minor underestimate of the IR luminosity (due to the exponential suppression of the rest-frame MIR emission). This effect however appears to be modest ($<15$\%) once we compare the IR luminosities adopted here with the ones that we would derive by using various dust templates from local IR galaxies \citep{silva98}. As we have no direct measurement of the dust SED in our sources yet, we opt for the simpler one-component model.}.} IR luminosities, $L_{\rm IR}$, are calculated by integrating equation~\ref{eq_mbb} between 3 and 1100 $\mu$m (rest frame; see \citealt{kennicutt12}). The far--IR luminosity $L_{\rm FIR}$ \citep[integrated between 42.5 and 122.5\,$\mu$m; see, e.g.,][]{helou88} is $L_{\rm FIR}=0.75\,L_{\rm IR}$ for our model. Given $T_{\rm dust}$ and $\beta$, we only need to normalize the continuum which we derive from the line--free channels at $\lambda_0\approx 158$\,$\mu$m. This gives a dust mass as well (see \citealt{venemans12,venemans16} and in prep.\ for further discussion). In order to ensure a consistent analysis, we also re-compute FIR luminosities for all the other quasars in the literature sample, starting from the published measurements of the continuum flux density at 158\,$\mu$m. 

Another useful quantity in this context is the \Cii{} (rest-frame) Equivalent Width (EW), defined as:
\begin{equation}\label{eq_ew}
\frac{\rm EW}{\rm km\,s^{-1}} = 1000\,\frac{F_{\rm line}\,{\rm [Jy\,km\,s^{-1}]}}{F_\nu({\rm cont})\,{\rm [mJy]}}.  
\end{equation}
The use of EWs has the practical advantage of circumventing any assumption of the shape of the dust SED. The inferred $L_{\rm [CII]}$, $L_{\rm FIR}$, and EW values for the targets in our sample are listed in Table~\ref{tab_size}.

Figure~\ref{fig_cii_fir}, {\em left} shows the \Cii{}/FIR luminosity ratio as a function of the FIR luminosity. This is a widely used diagnostic of ISM properties. The \Cii{}/FIR ratio is typically 0.003--0.01 in local, star-forming galaxies with modest dust temperature and luminosity, and drops by an order of magnitude towards the bright end (the so--called `\Cii{} deficit'), in particular in the presence of high--temperature, compact dust emission \citep{malhotra01,farrah13,diazsantos13,diazsantos14,diazsantos17,sargsyan14,herreracamus15}. The quasars in our sample span a wide range (over 1 dex) of \Cii{}/FIR values, from $\sim 0.0003$ as in local ULIRGs to $\sim 0.003$ as in local disk galaxies. This is consistent with what has been previously found for other smaller samples of $z>6$ quasars in the literature (e.g., \citealt{walter09,venemans12,venemans16,wang13,wang16,willott13,willott15,banados15,mazzucchelli17}). 

In Section~\ref{sec_CIIvsUV}, we discuss the dependence of $L_{\rm [CII]}$ on the rest-frame UV luminosity of the quasars.

\subsection{\Cii{} and FIR luminosity surface density}

An insight on the origin of the spread in the \Cii{}/FIR values is offered by the relatively tight relation between the \Cii{}/FIR and the FIR luminosity surface density, $\Sigma_{\rm FIR}$. We follow here the analysis presented by \citet{diazsantos17}, who studied fine-structure lines and dust emission in a sample of $\sim240$ IR--luminous galaxies at $z<0.1$ from the Great Observatories All-sky LIRG Survey (GOALS; \citealt{armus09}). We compute $\Sigma_{\rm FIR}=L_{\rm FIR}\,(2 \pi R_{\rm cont}^2)^{-1}$ based on the 2D Gaussian fit size of the continuum map (see Table~\ref{tab_size})\footnote{The factor 2 in the denominator accounts for the fact that $R_{\rm cont}$ roughly encompasses half of the total light; see a similar approach in, e.g., \citet{tacconi13}.}. We emphasize that our observations only marginally resolve the emission in most of the sources.
\citet{diazsantos17} show that the \Cii{}/FIR ratio in galaxies from the GOALS sample correlates with dust temperature (parametrized based on the ratio between the dust flux densities at 63\,$\mu$m and at 158\,$\mu$m) and the compactness of the IR continuum emission. Galaxies with more compact IR continuum emission tend to show higher dust temperature and lower \Cii{}/FIR ratio.

Our data generally follow the trend shown in \citet{diazsantos17} for local IR--luminous galaxies, with lower \Cii{} equivalent widths associated with higher FIR surface densities. This might also reflect a diversity in the dust temperature in quasar host galaxies, with increasing dust temperature at decreasing \Cii{}/FIR or \Cii{} EW; however, the scatter is significant, and the present data do not support a straightforward derivation of $T_{\rm dust}$ from either $\Sigma_{\rm FIR}$ or the \Cii{}/FIR ratio.

We can also infer the \Cii{} luminosity surface density, $\Sigma_{\rm [CII]}$, using $R_{\rm [CII]}$, i.e., the half-size of the beam--deconvolved \Cii{} emission. The resulting values are listed in Table~\ref{tab_size}.

\begin{figure}
\begin{center}
\includegraphics[width=0.99\columnwidth]{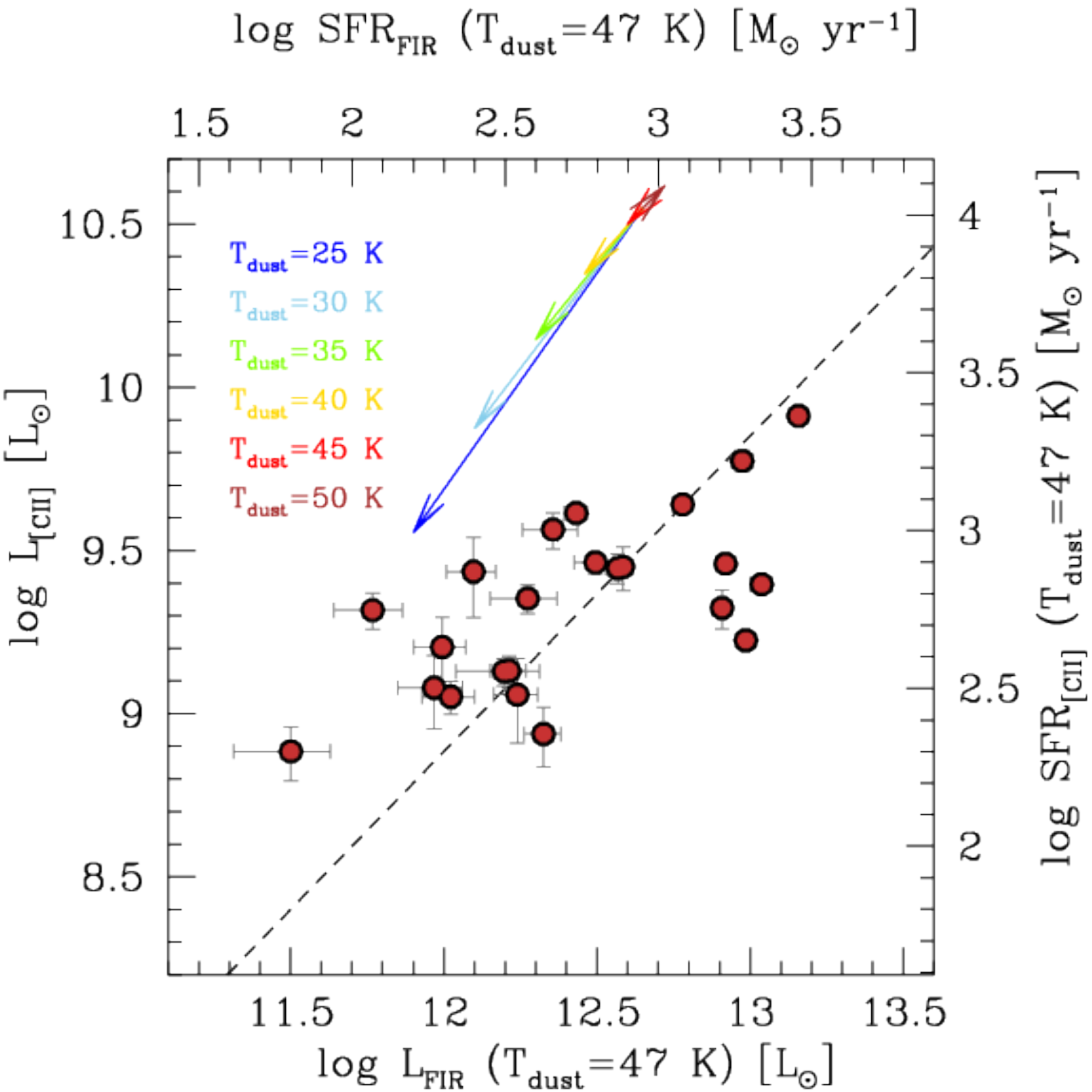}\\
\end{center}
\caption{Comparison between the star-formation rate estimates derived from the \Cii{} luminosity (following \citealt{herreracamus15}) and from the dust continuum luminosity (following \citealt{kennicutt12}), in the assumption of $T_{\rm dust}=47$\,K. The dashed line is the 1--to--1 case. The quasar host galaxies in our sample cover a range of about 1.5 dex in star formation rate. The two prescriptions for SFR estimates are correlated, although substantial scatter is present. The arrows show how the SFR estimates would change if we adopt different dust temperatures.}
\label{fig_sfr_sfr}
\end{figure}

\begin{figure}
\begin{center}
\includegraphics[width=0.99\columnwidth]{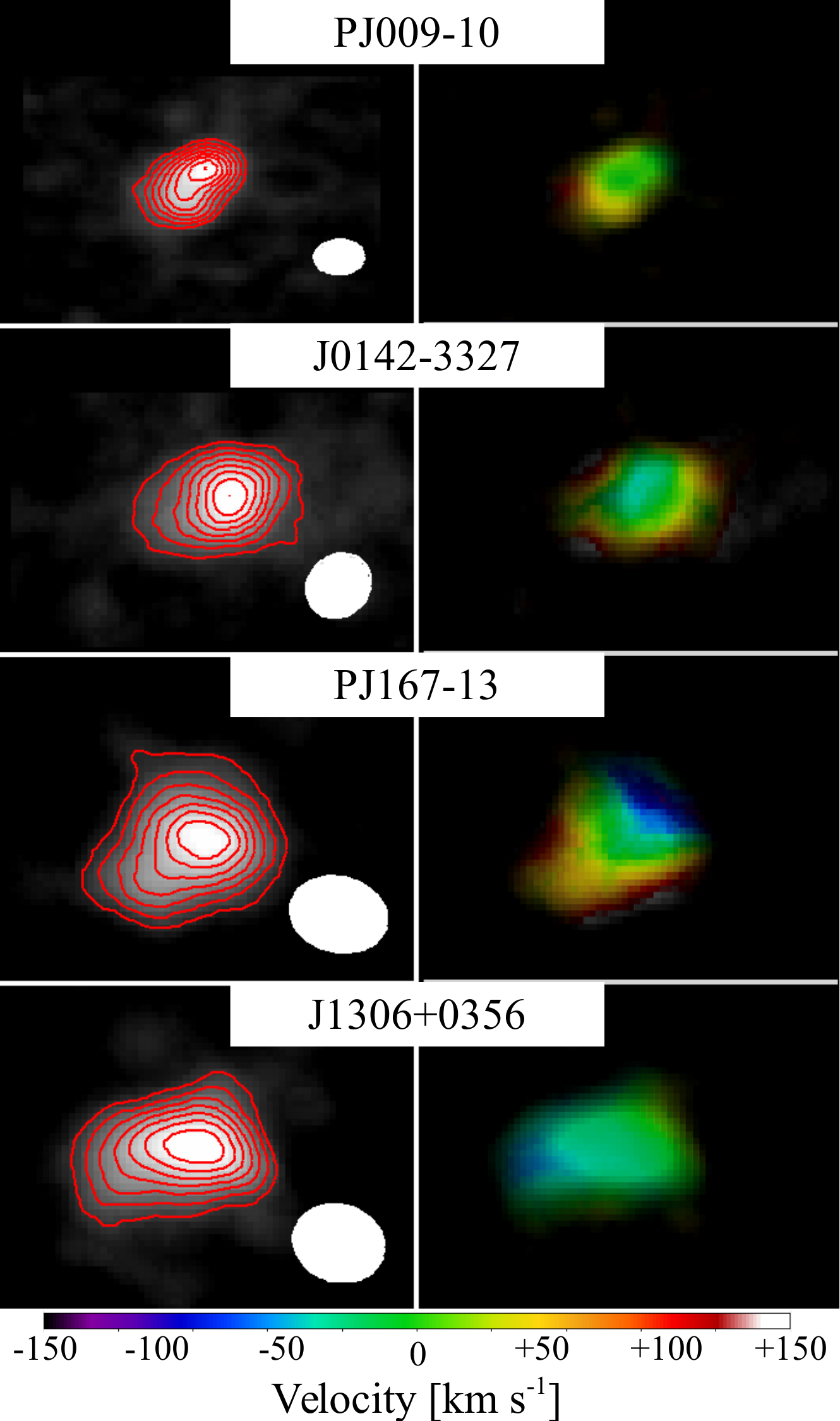}\\
\end{center}
\caption{Moment zero ({\em left}) and one ({\em right}) maps of the four objects in our sample with S/N$>$10 in the 2D gaussian fit of the line, and with clearly resolved emission (observed size $>1.4\times$ the beam, see Figure~\ref{fig_beam_ima_spc}). Here the maps have been re-imaged using robustness parameter = 0.5 to increase the angular resolution. In the moment zero maps, the contours mark the 30\%, 40\%, 50\%, \ldots 90\% of the peak emission. The beam of the observations is plotted in each panel for reference as a white ellipse. Each panel is $4.5''\times3.5''$, corresponding to $\sim 25$\,kpc\,$\times20$\,kpc at the redshift of these quasars. North is up, East to the left. The velocity maps have all the same color scale, centered at the bulk of the \Cii{} emission. Clear velocity gradients are observed in PJ167-13 and J1306+0356, and to a minor extent also in PJ009-10 and J0142-3327.}
\label{fig_vel_maps}
\end{figure}

\subsection{Star formation rates}

Both the \Cii{} luminosity and the dust continuum luminosity have been used in the literature to infer the star-formation rates (SFRs) of galaxies. Observational advantages in using \Cii{} are that 1) the line is bright and therefore easy to detect in a single frequency setting if the redshift of the source is known within a few thousand \kms{}, 2) it is ubiquitously found in galaxies, and 3) it is rarely affected by saturation or absorption, except in the densest starburst nuclei. However, the ``deficit'' of \Cii{} emission in the brightest sources as well as its dependence on metallicity challenge its applicability as a SFR tracer (in particular for very luminous and compact starbursts, and for very metal--poor galaxies; see, e.g., \citealt{delooze14,herreracamus15}). Conversely, the photometric sampling of the dust continuum emission enables SFR estimates even in the absence of precise redshift information (see \citealt{kennicutt12} for a review on this topic). However, in order to effectively pin down the integrated dust emission, it is important to sample the dust SED close to its peak (rest-frame wavelengths $\lambda \lsim 100$\,$\mu$m), which is often challenging from the ground (as the frequencies of interest typically are $<500$\,GHz, where the atmospheric transmission is limited). 

Our \Cii{}--based SFR estimates follow the fit to the \Cii{}--SFR (24\,$\mu$m--based) relation in 46 local galaxies from the KINGFISH sample \citep{kennicutt11} derived by \citet{herreracamus15}: 
\begin{equation}\label{eq_cii_sfr}
\frac{\rm SFR_{[CII]}}{\rm M_\odot\,yr^{-1}}=0.052\,\left(\frac{L_{\rm [CII]}\,\Psi(y)}{\rm 10^{40}\,erg\,s^{-1}}\right)^{1.034}
\end{equation}
where $\Psi(y)=(y/y_{\rm t})^\alpha$ is a correction term dependent on dust temperature, with $y=\nu F_\nu$(70\,$\mu$m)/$\nu F_\nu$(160\,$\mu$m), $y_{\rm t}$=1.12, and $\alpha=1.2$. Adopting the calibration by \citet{delooze14} results in comparable SFR estimates.

For the dust--based SFR estimates, we follow \citet{kennicutt12}:
\begin{equation}\label{eq_ir_sfr}
\frac{\rm SFR_{IR}}{\rm M_\odot\,yr^{-1}}=1.49\times10^{-10}\,\frac{L_{\rm IR}}{\rm L_\odot}
\end{equation}

In Figure~\ref{fig_sfr_sfr}, we compare the \Cii{} and FIR luminosities and the corresponding SFRs, computed assuming $T_{\rm dust}=47$\,K.  The two estimates of SFR span approximately 1.5\,dex and appear to correlate, although with substantial scatter. Different assumptions on the dust temperatures would affect both estimates of SFR in a similar way, leading to lower SFR values for lower dust temperatures. 

The inferred SFR surface densities (the quantities in Figure~\ref{fig_sfr_sfr} divided by our coarse estimates of the size of the emitting region) are always $<100$\,\Msun{}\,yr$^{-1}$\,kpc$^{-2}$, i.e., well below the theoretical Eddington limit ($\sim1000$\,\Msun{}\,yr$^{-1}$\,kpc$^{-2}$; see, e.g., \citealt{scoville04}, \citealt{thompson05}, \citealt{walter09}).

\begin{figure}
\begin{center}
\includegraphics[width=0.99\columnwidth]{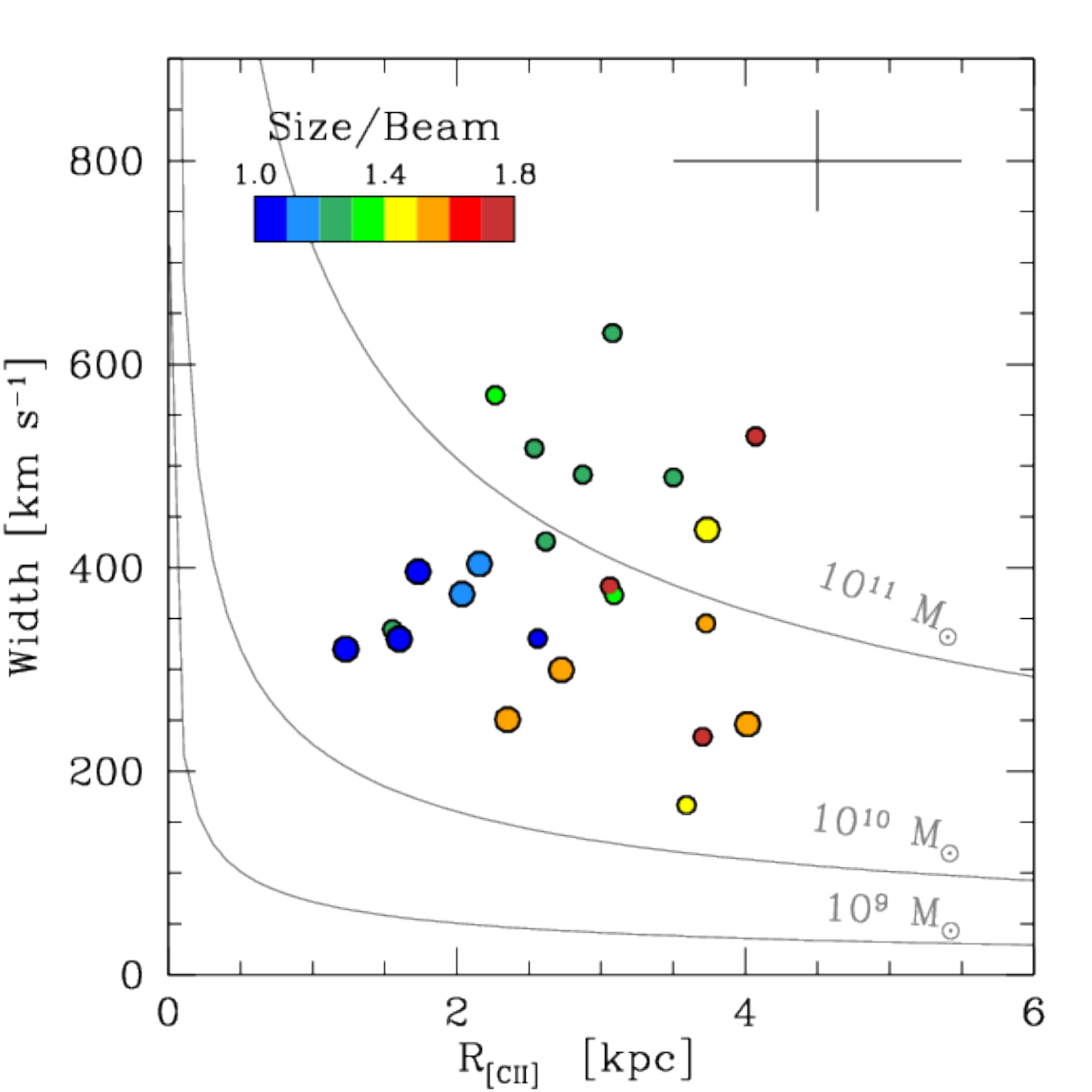}\\
\end{center}
\caption{Line width (plotted as full width at half maximum) as a function of the radius of the \Cii{}--emitting region ($R_{\rm [CII]}$) in the quasars in our sample. The color code is the same as in Figure~\ref{fig_beam_ima_spc}. Sources with S/N$>$10 in the \Cii{} map are highlighted with bigger symbols. Typical error bars are shown in the top--right corner. Under the assumption of rotation--dominated dynamics (see equation \ref{eq_mdyn2}), the combination of these quantities yields an order-of-magnitude estimate of the dynamical mass enclosed within $R_{\rm [CII]}$, as shown by the loci of constant mass. We find dynamical masses of $2\times10^{10}-2\times10^{11}$\,\Msun{}. All but one of the high--S/N sources have $M_{\rm dyn}\approx 4\times10^{10}$\,\Msun{}.}
\label{fig_mdyn}
\end{figure}

\begin{figure}
\begin{center}
\includegraphics[width=0.99\columnwidth]{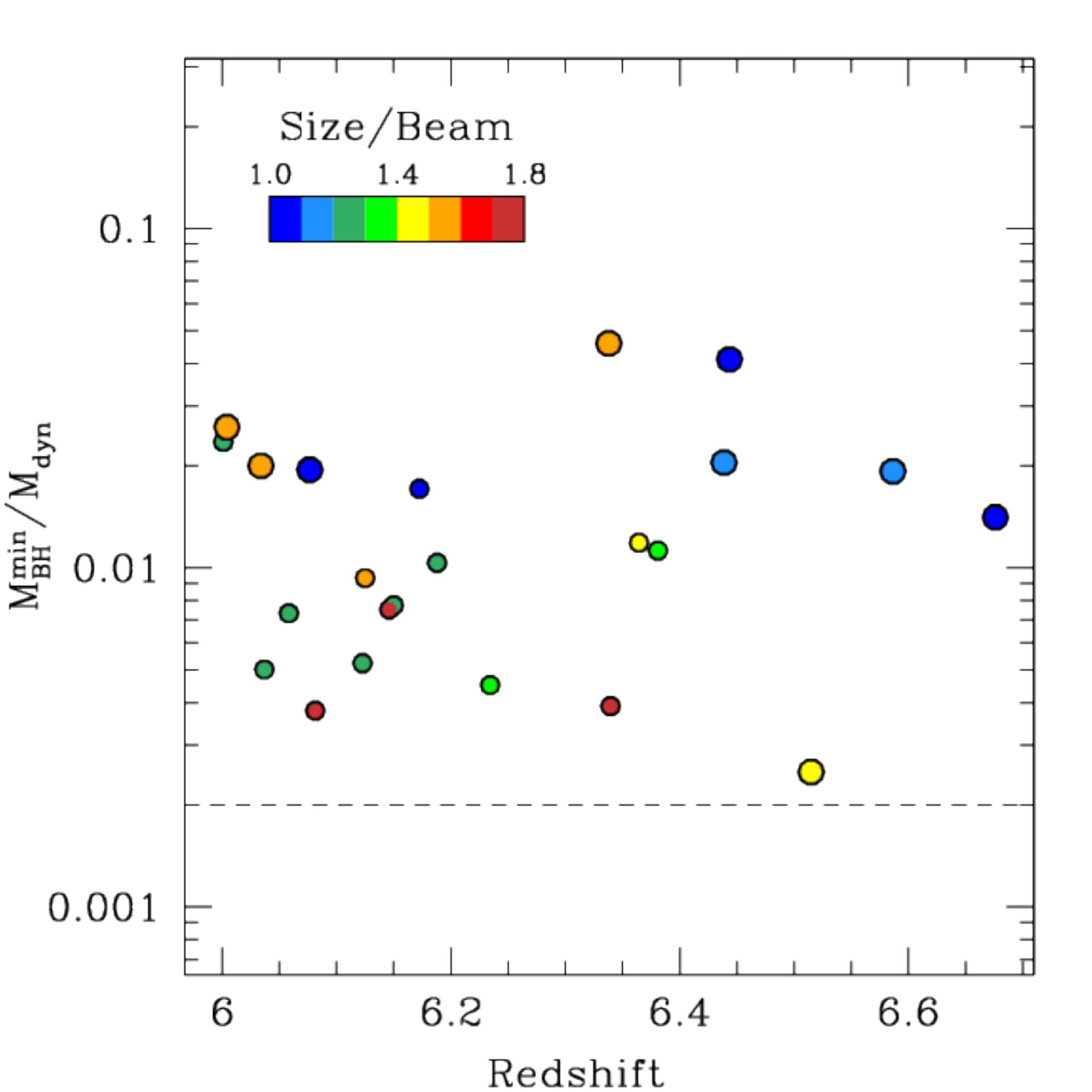}\\
\end{center}
\caption{Constraints on the black hole -- host galaxy mass ratio for the main sample of this work, as a function of the \Cii{} redshift. The color code is the same as in Figs.~\ref{fig_beam_ima_spc} and \ref{fig_mdyn}. Objects with S/N$>$10 in the \Cii{} map are highlighted with larger symbols. The minimum black hole mass $M_{\rm BH}^{min}$ is computed from the rest-frame UV continuum luminosity, by assuming that the quasars are emitting at Eddington luminosity. The dynamical mass is derived via equation \ref{eq_mdyn2}, and might be considered as an upper limit for the marginally--resolved sources (blue points) or if the dynamics is dispersion supported (see equation \ref{eq_mdyn}). The plotted ratio can therefore be considered as a lower limit. The ratio observed in local galaxies is marked with a dashed line. All our quasars clearly lie above the local value. In particular, all but one of the high--S/N sources have $M_{\rm BH}^{\rm min}/M_{\rm dyn}\approx0.03$, i.e., 1 dex above the local value.}
\label{fig_mbh_mhost}
\end{figure}

\subsection{Velocity maps and dynamical masses}\label{sec_mdyn}

Four sources in our sample are well--resolved in \Cii{} (i.e., the observed size of the continuum--subtracted \Cii{} emission from the 2D gaussian modeling is $>1.4\times$ the beam): PJ009--10, J0142--3327, PJ167--13, and J1306+0356 (see Figure~\ref{fig_beam_ima_spc}). In Figure~\ref{fig_vel_maps}, we show the zero moment and the velocity field maps of these four sources. These are created after re-imaging the cubes using robustness parameter = 0.5, which increases the relative weights of the long baselines, thus improving the angular resolution of the imaged cubes. PJ009--10 shows an elongated morphology along the North-West -- South-East axis. Its velocity map shows a small velocity gradient. Similarly, J0142--3327 appears extended along the East--West direction, but the velocity structure is less clear. On the other hand, PJ167--13 shows a clearly resolved velocity gradient from North-West towards South-East, with a peak--to--peak velocity difference exceeding 400\,\kms{} along the line of sight. J1306+0356 shows a velocity gradient along the East--West direction, although in this case the peak--to--peak velocity difference is lower ($\sim 200$\,\kms{}). The resolution of the available data is insufficient to assess whether the kinematics of the \Cii{}--emitting gas in the host galaxies of these two quasars are dominated by ordered rotation or if the underlying velocity structure is more complex (see, for instance, the high-resolution studies of $z>6$ quasars presented in \citealt{venemans17a} and \citealt{shao17}). In particular, given the present data quality it is impossible to rule out whether part of the spatially--resolved \Cii{} emission is associated with a close satellite galaxy of the quasar host galaxy, similar to the cases discussed in \citet{decarli17}.

We can make rough estimates of the host galaxy dynamical masses from our observations. The dynamical mass in a dispersion--dominated system can be expressed as:
\begin{equation}\label{eq_mdyn}
M_{\rm dyn}= \frac{3}{2} \, \frac{R_{\rm [CII]} \sigma_{\rm line}^2}{G},
\end{equation}
where $R_{\rm [CII]}$ is the radius of the \Cii{}--emitting region (defined as the major semiaxis of the 2D Gaussian fit of the \Cii{} map), $\sigma_{\rm line}$ is the line width from the gaussian fit of the \Cii{} spectra, and $G$ is the gravitational constant. If the line width is dominated by rotation, the gas appears as a flat disk with an inclination angle $i$ \citep[see, e.g.,][]{wang13,willott15}. In this case:
\begin{equation}\label{eq_mdyn2}
M_{\rm dyn}= G^{-1}\,R_{\rm [CII]}\,(0.75\,{\rm FWHM} /\sin i)^2.
\end{equation}
Here, 0.75 is a factor to scale the line FWHM to the width of the line at 20\% of the peak, in the case of a Gaussian profile, following \citet{willott15}. If we assume an inclination of $i=55^\circ$ \citep[following][who derived it as the median inclination angle from the Wang et al.~(2013) sample]{willott15}, the dynamical mass inferred with equation \ref{eq_mdyn2} is $3.1\times$ larger than the one estimated with equation \ref{eq_mdyn} (see \citealt{deblok14} for a detailed discussion on deriving dynamical mass constraints from unresolved observations). 

In Figure~\ref{fig_mdyn} we show the \Cii{} line width and size for the quasar host galaxies in our sample. These are comparable with the ones reported in the literature for $z>6$ quasar host galaxies \citep{walter04,wang13,willott15,venemans17a}. The combination of size and line width implies that the targeted host galaxies have dynamical masses in the range $2\times 10^{10}-2\times10^{11}$\,\Msun{}, if we adopt equation \ref{eq_mdyn2}. In particular, all but one of the sources detected with S/N$>$10 in the \Cii{} map have $M_{\rm dyn}\approx4\times 10^{10}$\,\Msun{}. We stress however that, given the limited angular resolution of our observations, the dynamical mass estimates in some of the sources in our sample might be overestimated.

\subsection{Black hole to host galaxy mass ratio}\label{sec_mbh_mhost}

In the local universe, the mass of black holes in galaxy nuclei correlates with the host galaxy stellar mass (as well as with other large-scale properties of the galaxy, such as the stellar velocity dispersion). The typical mass ratio is $M_{\rm BH}/M_{\rm host} \sim 0.002$ (e.g., \citealt{marconi03,haering04,sani11,kormendy13}; a factor $\sim10$ lower according to \citealt{reines15}). Whether this ratio evolves with redshift is a matter of debate. Observations of the host galaxy starlight in conditions of natural seeing \citep[e.g.,][]{decarli10,targett12,matsuoka14}, using adaptive optics \citep[e.g.,][]{falomo05,inskip11}, or capitalizing on the exquisite angular resolution of the {\em Hubble Space Telescope} \citep[e.g.,][]{dunlop03,bennert11,schramm13,park15}, in some cases aided by natural magnification \citep{peng06,ding17} point towards a higher black hole to host galaxy mass ratio in quasars at redshift $z=1$--$4$, compared to local relations \citep[although some studies, e.g.,][found no evidence for an evolution in the black hole to host galaxy mass ratio]{jahnke09,cisternas11}. Studies exploiting spatially--unresolved observations of the spectral energy distribution of fainter active galactic nuclei \citep[e.g.,][]{merloni10} also suggest that black holes were `overmassive' at high redshift compared with galaxies of the same stellar mass in the local universe. Spatially--resolved observations at mm and radio wavelengths of gas in the host galaxy of high--redshift quasars have enabled dynamical estimates of the host galaxies up to the highest redshifts \citep[e.g.,][]{walter04,schumacher12,wang13,willott13,willott15,venemans16,venemans17a,shao17}. They all consistently find a tendency towards a higher $M_{\rm BH}/M_{\rm dyn}$ ratio in high redshift quasars than the value observed in local galaxies, with up to a factor $\sim10$ discrepancy at $z>6$. This general consensus in the observations might however be undermined by selection biases: since high--redshift studies focus on luminous quasars, they might privilege galaxies hosting more massive black holes (which can reach higher luminosities, as their Eddington luminosity is also higher) than the average population. The works by \citet{lauer07}, \citet{decarli10}, \citet{schulze14}, \citet{degraf15}, \citet{shankar16}, and \citet{volonteri16} extensively discuss these issues.

At $z>6$, black hole masses are usually inferred from spectroscopic observations of the Mg{\sc ii} broad emission line at 2796\,\AA{}, which is shifted in the NIR $K$ band \citep[e.g.][]{willott03,shen13,derosa11,derosa14,mazzucchelli17}. The typical black hole mass is $\sim 10^9$\,\Msun{}, implying that these quasars radiate close to their Eddington limit \citep{jiang07,derosa11,derosa14,mazzucchelli17}. Sensitive NIR spectroscopy is not available for all the quasars in our sample yet (the results from our dedicated program will appear in Farina et al.~in prep.). However, we can set tentative constraints on the black hole masses by requiring that their luminosity is lower than the Eddington luminosity. This allows us to infer a limit on the minimum black hole mass of the quasars in our sample:
\begin{equation}\label{eq_edd}
\frac{M_{\rm BH}^{\rm min}}{\rm M_\odot} = \frac{L_{\rm bol}}{1.26\times10^{38}\,{\rm erg\,s^{-1}}}.
\end{equation}
We derive the bolometric luminosity from the observed flux density at 1450\,\AA{}, following the conversion derived by \citet{runnoe12} and recomputed in \citet{venemans16}: 
\begin{equation}\label{eq_bol}
\log \frac{L_{\rm bol}}{\rm erg\,s^{-1}}=4.553+0.911\,\log \frac{\lambda L_{\lambda} (1450\,\AA)}{\rm erg\,s^{-1}}
\end{equation}
We derive lower limits on the black hole masses of $M_{\rm BH}^{\rm min}$=$(0.3-3)\times 10^9$\,\Msun{}. In Figure~\ref{fig_mbh_mhost} we compare these black hole mass limits with the host galaxy dynamical masses in the conservative case of rotationally--supported gas dynamics (equation \ref{eq_mdyn2}). We find that the mass ratio between black holes and host galaxies is well above the value observed in the local universe. In particular, all but one of the sources detected at high S/N in \Cii{} have a mass ratio of $>$0.03, i.e., $>$15 times higher than the value observed in the local universe ($>$120 times higher than the expected value using the local relations in \citealt{reines15}). Using an isothermal model for the host galaxy dynamics instead (equation \ref{eq_mdyn}) would lead to even higher $M_{\rm BH}^{\rm min}/M_{\rm dyn}$ ratios.

\begin{figure*}
\begin{center}
\includegraphics[width=0.99\columnwidth]{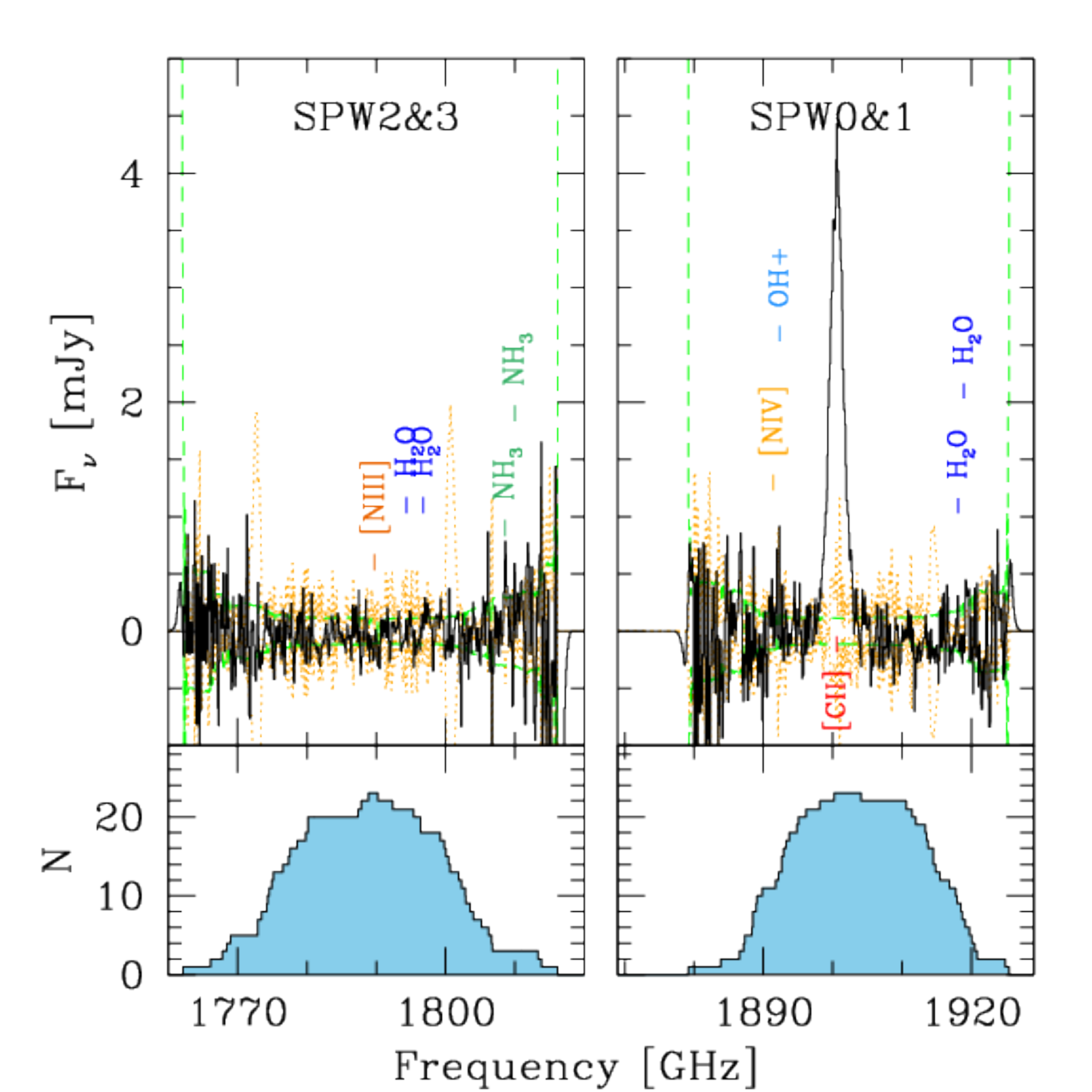}
\includegraphics[width=0.99\columnwidth]{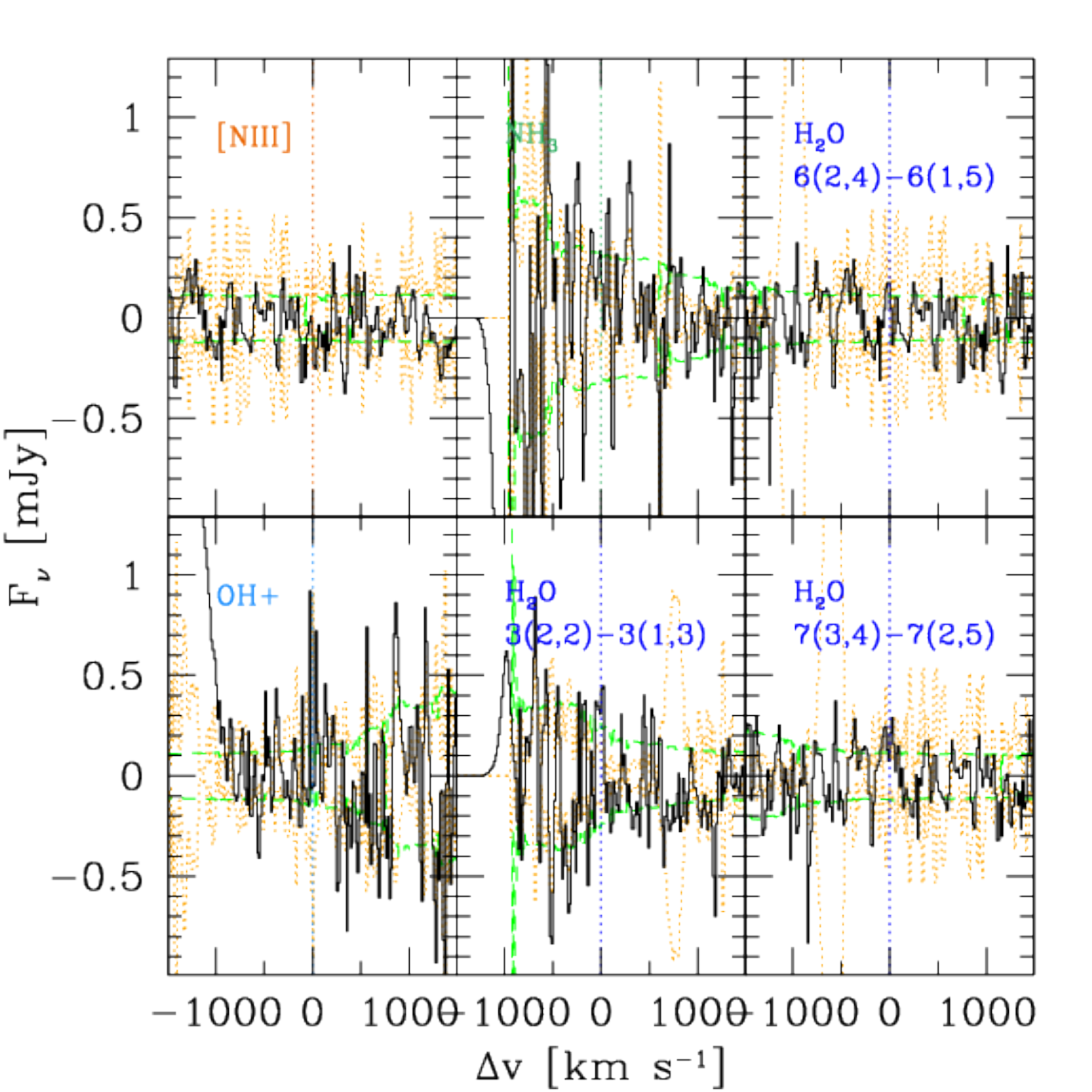}\\
\includegraphics[width=0.99\columnwidth]{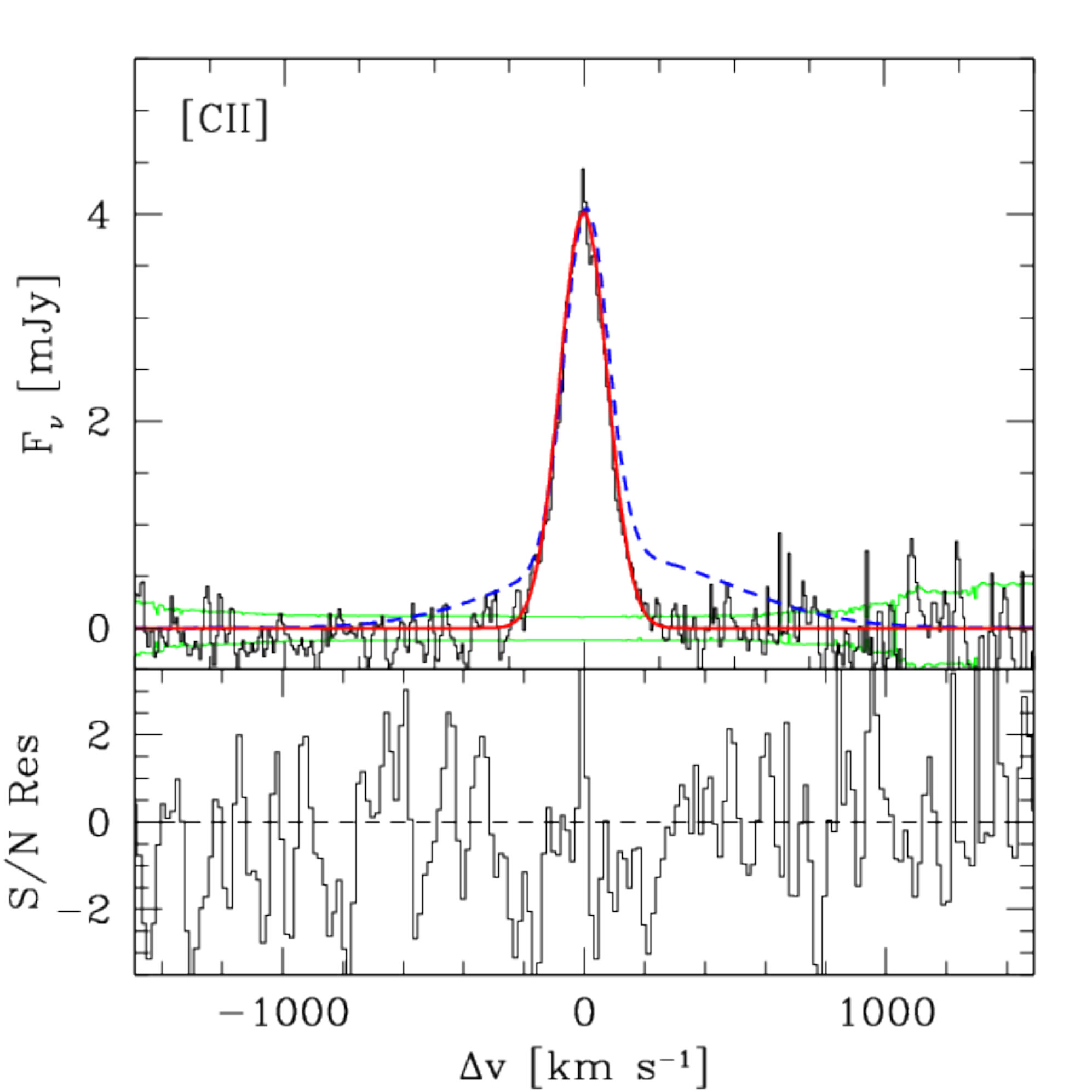}
\includegraphics[width=0.99\columnwidth]{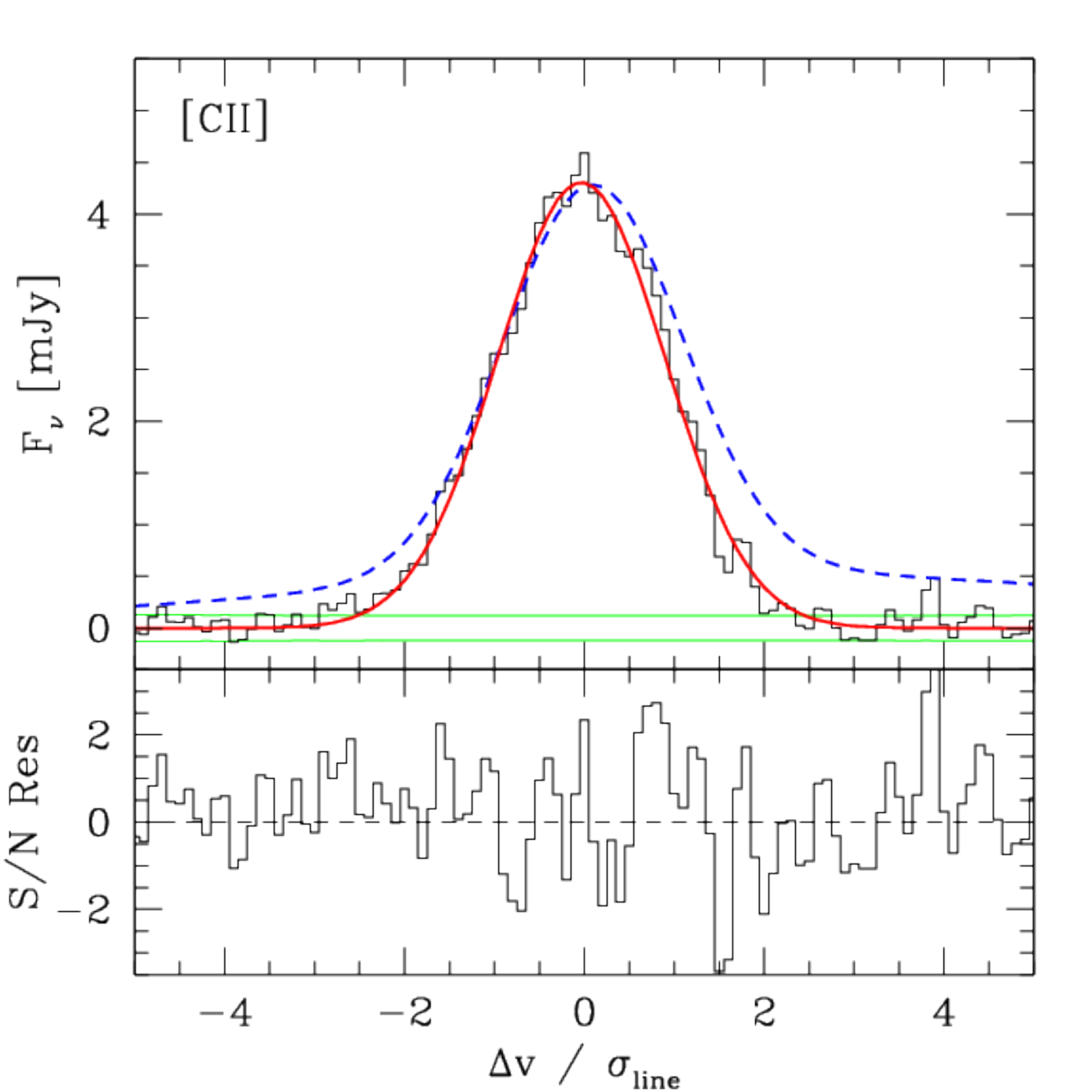}\\
\end{center}
\caption{Stacked spectra at 1900\,GHz of the \Cii{}--detected quasars in our sample. We plot the stacked spectra as black histograms, their uncertainties (computed based on the noise of individual spectra) as green histograms, and the standard deviation (computed based on the variance between individual spectra) as orange, dotted histograms. {\em Top left:} Weighted--average stack of the individual spectra, shifted to rest frame based on the \Cii{} redshift. Other lines that fall in the frequency range are marked. The bottom panels show the number of spectra used in the stack as a function of frequency.  {\em Top right:} Zoom-in on the stacked spectrum of the quasar host galaxies in our sample, highlighting the expected frequencies of a number of other lines for which we have coverage. No detection is found for any of these lines. {\em Bottom left:} Weighted--average stack of individual \Cii{} spectra, highlighting the \Cii{} line. The best--fit Gaussian model of the stacked line is shown as a thick red line. For comparison, the \Cii{} line profile of J1148+5251 as modeled in \citet{cicone15}, showing a prominent outflow feature, is shown with a dashed blue line. The fit of J1148+5251 is normalized to match the peak flux density of the stacked \Cii{} line in our sample. The bottom panel shows the residual from the fit, normalized by the noise per pixel. The stacked spectrum does not reveal any significant deviation from a Gaussian profile. {\em Bottom right:} Same as on the bottom--left panel, but this time stacking the spectra scaled by the width of each line. Also in this case, no significant deviation from Gaussianity is reported.  }
\label{fig_spc_templ}
\end{figure*}

\subsection{Stacked spectra}

Figure~\ref{fig_spc_templ} shows the stacked spectra of the quasars in our survey. We include only the \Cii{}--detected sources as no other redshift indicator is precise enough for stacking (see Figure~\ref{fig_dv}). After shifting all the observed spectra to the rest frame, we subtract the continuum (from the spectral fit), and average by weighting by the inverse of the variance (from the observed spectral noise). The uncertainty on the composite spectrum is obtained from the inverse square root of the sum of the weights. The resulting spectrum allows us to search for other faint lines in the rest-frame range 1775--1805\,GHz (from SPW2\&3) and 1890--1920\,GHz (from SPW0\&1). This range encompasses the fine structure lines of [N{\sc iii}] 1789.806\,GHz and [N{\sc iv}] 1891.435\,GHz; ammonia NH$_3$ lines at 1808.93 and 1810.38\,GHz; the water lines H$_2$O 6(2,4)-6(1,5) at 1794.789\,GHz, 7(3,4)-7(2,5) at 1797.159\,GHz, 5(2,3)-4(3,2) at 1918.485\,GHz, and 3(2,2)-3(1,3) at 1919.360\,GHz; as well as a multiplet of OH+ lines at 1892.0--1892.2\,GHz (blended with the [N{\sc iv}] line). None of these transitions is significantly detected in our stacked spectrum. By integrating within $\pm 400$\,\kms{} around the nominal frequency of the lines, we obtain limits on the integrated flux of these lines in emission compared with the \Cii{} line (see Table~\ref{tab_otherlines}). We also do not detect these lines if we stack the continuum--normalized spectra (which allows us to search for the same lines in absorption). Similarly, by integrating the template based on continuum--normalized spectra within $\pm 400$\,\kms{} of the expected frequency of the lines, we infer limits on the equivalent widths of these lines (see Table~\ref{tab_otherlines}). These limits are consistent with the few direct measurements and limits available for these transitions in galaxies in the literature. For example, in NGC\,4418 and Arp\,220 the H$_2$O 3(2,2)-3(1,3) line has EWs of $9.2\pm2.3$\,\kms{} and $14.5\pm0.9$\,\kms{}, respectively \citep{gonzalezalfonso12}. In HFLS-3 ($z=6.34$), all the water transitions discussed in our study are undetected, with limits on the EW of $<$1630\,\kms{} \citep{riechers13}. 

\begin{table}
\caption{{\rm Limits on the strength of secondary lines covered in the stacked quasar spectra shown in Figure~\ref{fig_spc_templ}. (1) Transition ($X$). (2) Rest-frame frequency. (3) 5-$\sigma$ limit on the \Cii{}/$X$ luminosity ratio. (4) 5-$\sigma$ limit on the equivalent width of transition $X$.}} \label{tab_otherlines}
\begin{center}
\begin{tabular}{cccc}
\hline
    $X$                 & $\nu_0$          & \Cii{}/$X$ & EW($X$)    \\
                        & [GHz]            &            & [\kms]     \\
    (1)                 &  (2)             &   (3)      &  (4)       \\
    \hline
[N{\sc iii}]            & 1789.806         & $>27$      & $<44$      \\
NH$_3$                  & 1808.93, 1810.38 & $>9$       & $<143$     \\
OH+                     & 1892.0-1892.2    & $>19$      & $<45$      \\
H$_2$O$_{3(2,2)-3(1,3)}$& 1919.360         & $>10$      & $<69$      \\
H$_2$O$_{5(2,3)-4(3,2)}$& 1918.485         & $>10$      & $<51$      \\
H$_2$O$_{6(2,4)-6(1,5)}$& 1794.789         & $>25$      & $<57$      \\
H$_2$O$_{7(3,4)-7(2,5)}$& 1797.159         & $>25$      & $<57$      \\
\hline
\end{tabular}
\end{center}
\end{table}

We also create a second spectral template by scaling the velocity axis, so that all the \Cii{} lines have the same width. We do so by scaling the abscissae of the individual spectra by the best-fit width of the \Cii{} line $\sigma_{\rm line}$ (see Figure~\ref{fig_spc_templ}, {\em bottom right}). 

The stacked spectra show no evidence of deviations from a Gaussian curve. We follow \citet{deblok14} in order to put this result in the context of the geometry and kinematics of the \Cii{}--emitting region. If the gas dynamics are supported by rotation in a disk, with the rotational velocity steeply increasing at small radii, and then flattening out at large radii (as seen in local spiral galaxies), we would naively expect a double--horned line profile, as observed in unresolved H{\sc i} observations of disk galaxies \citep[e.g.,][]{catinella10}. Its absence can be explained with the following arguments: 1) The gas is turbulent, i.e., the dispersion velocity term is at least comparable to the rotational velocity component along the line of sight; 2) The emission in the central beam of our observations captures a scale that is comparable with or smaller than the rising part of the velocity curve (i.e., $R_{\rm [CII]}\lsim h$, where $h$ is the scale length of the exponential disk, following the parametrization in \citealt{deblok14}). These explanations are not mutually exclusive. High angular resolution observations of a few $z>6$ quasars revealed very compact \Cii{} emission \citep[$\sim1$\,kpc; see][]{walter09,venemans17a,shao17}, which favors the second scenario. Even when the \Cii{} emission is clearly extended (e.g., J0305--3150, see \citealt{venemans16}; or P167--13 presented here), the luminosity--weighted size estimates available so far are in the $3-5$\,kpc range, and extend only to a couple of beam radii. Observations with both higher angular resolutions and higher surface brightness sensitivity on large samples of quasars are needed in order to expose the faint, diffuse \Cii{} emission on the outskirt of individual quasar host galaxies, thus allowing us to accurately constrain the global dynamical properties of these systems.

The lack of non-Gaussian components in the stacked \Cii{} line profile has interesting implications in the search for outflows in the host galaxies of these quasars. J1148+5251 \citep{cicone12,cicone15} and Mrk231 \citep{feruglio15} are remarkable examples of IR--bright quasars with non-Gaussian line profiles. In particular, \citet{cicone15} model the \Cii{} line profile in J1148+5251 as the sum of narrow and broad components (see Figure~\ref{fig_spc_templ}). If such a feature were common in the quasars in our sample, we should clearly detect it in the stacked spectra. However, this is not the case, suggesting that J1148+5251 is unique compared to the bulk of the $z>6$ quasar population. No significant deviation from the Gaussian profile is detected in any of the individual spectra either (see Figure~\ref{fig_alma_spc}), with the caveat that some of the spectra only have a relatively modest S/N for this kind of analysis.

\begin{figure*}
\begin{center}
\includegraphics[width=0.99\columnwidth]{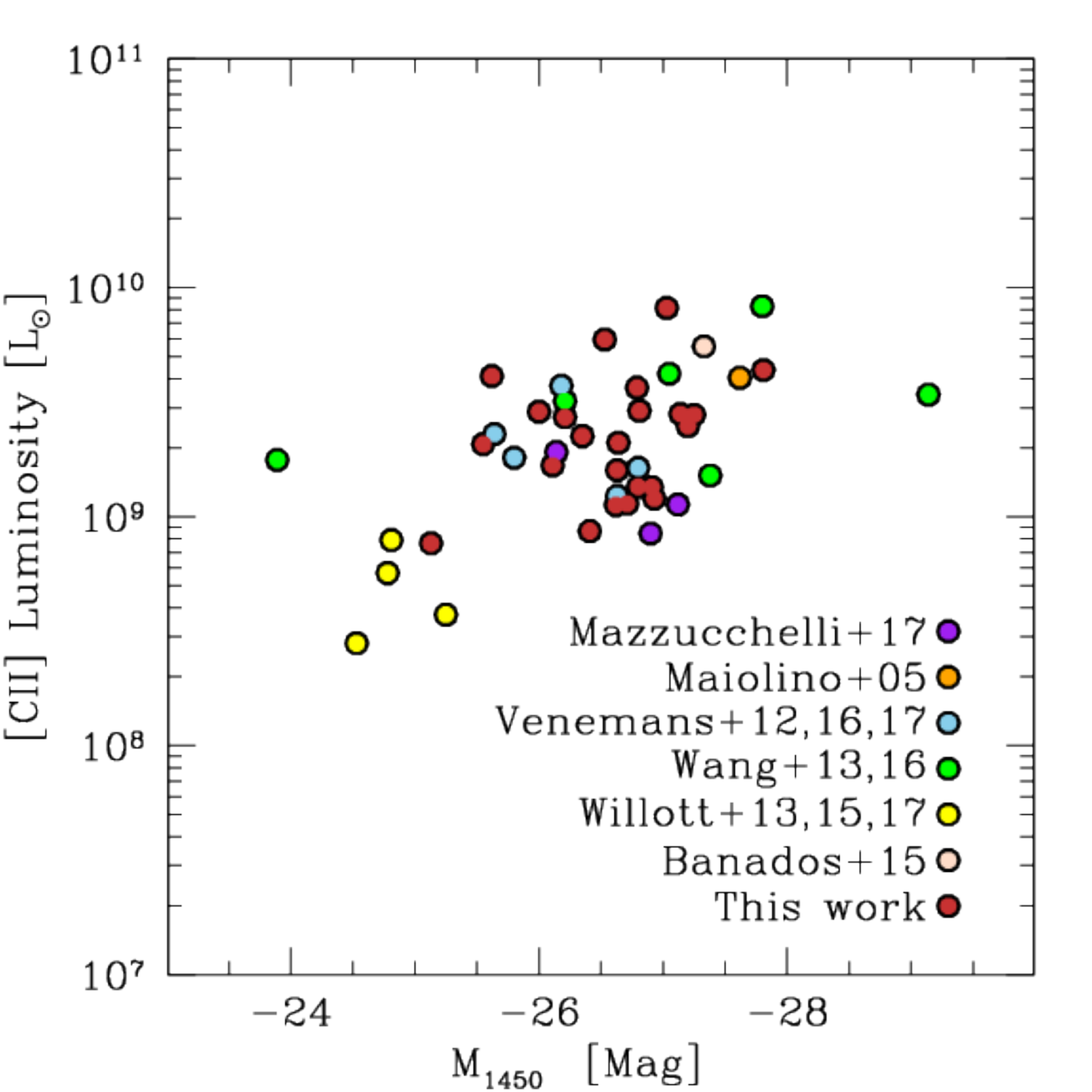}
\includegraphics[width=0.99\columnwidth]{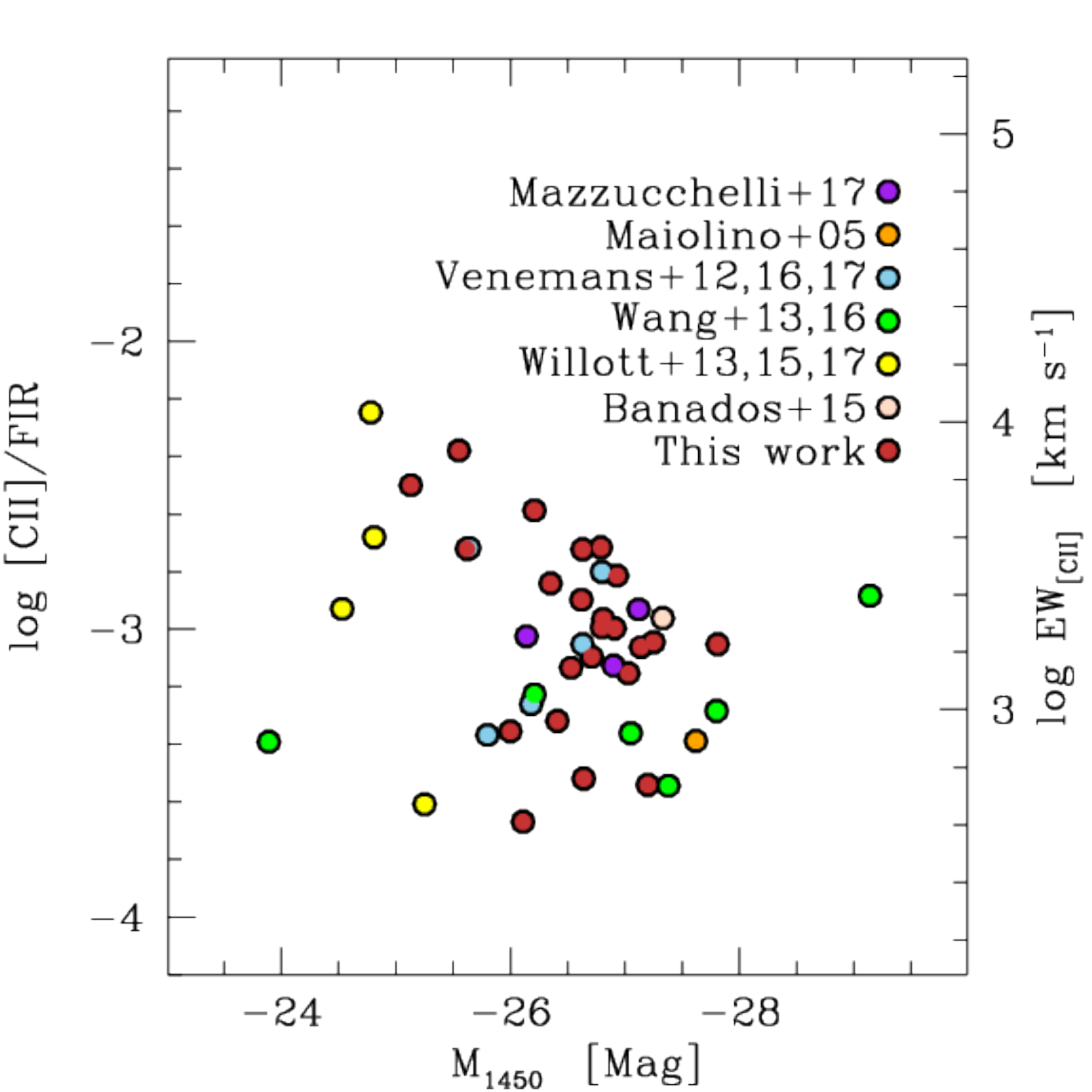}\\
\end{center}
\caption{The dependence of the \Cii{} luminosity and of the \Cii{}/FIR ratio (or the \Cii{} equivalent width) on the quasar UV luminosity, reported as absolute magnitude at 1450\,\AA{} in the rest frame. A weak correlation between UV and \Cii{} luminosity is observed, although this is mostly driven by the inclusion of the \citet{willott13,willott15,willott17} sample at lower luminosities. The \Cii{} equivalent width and \Cii{}/FIR ratio show no correlation with the quasar UV luminosity.}
\label{fig_Muv}
\end{figure*}

\subsection{\Cii{} dependence on quasar UV luminosity}\label{sec_CIIvsUV}

In Figure~\ref{fig_Muv} we find a mild dependence of the \Cii{} luminosity in our sample on the quasar UV luminosity, expressed in terms of the absolute magnitude at 1450\,\AA{} rest frame, $M_{1450}$. However, this conclusion is mostly driven by the contribution of the \citet{willott13,willott15} sample, that targets significantly lower UV and \Cii{} luminosities. On the other hand, the \Cii{} equivalent width (right panel) is practically independent of the quasar UV luminosity.

The lack of a correlation between the \Cii{} equivalent width and the quasar UV luminosity may provide us with clues about the physical mechanisms responsible for heating the dust and for exciting the carbon ions. To first order, the more UV--luminous is the quasar, the higher are the dust temperature and the degree of carbon ionization (beyond the single ionization associated with the \Cii{} line targeted here), thus we expect lower \Cii{}/FIR luminosity ratio. From an empirical point of view, however, such a trend may break down if: 1) the observed UV emission of the quasar is (at least in some objects) affected by dust reddening along the line of sight; 2) the UV emission from the quasar boosts (instead of suppresses) the \Cii{} emission in regions where the column densities are sufficient to shield photons with energy $E_\gamma>24.4$ eV (responsible for the second ionization of carbon); 3) the ISM in the host galaxy is multi-phase, in which a more extended and diffuse \Cii{}--emitting region powered by star formation is superimposed on the central region which is strongly affected by the quasar radiation. This latter scenario is supported by the high \Cii{}/[C{\sc i}]$_{2-1}$ ratios observed in some $z>6.5$ quasars, which are in tension with an X-ray--driven (i.e., AGN--powered) excitation of the gas \citep{venemans17b}.

\section{Discussion \& Conclusions}\label{sec_discuss}

We presented a systematic study of the \Cii{} line emission in 27 quasars at $z\gsim6$ with ALMA. This effort more than doubles the number of $z\gsim6$ quasars observed in \Cii{} to date. Our main findings are:
\begin{itemize}
\item[{\em i-}] We detect \Cii{} (and the dust underlying continuum) in 23 of the 27 targeted quasars (detection rate: $\sim 85$\%) with \Cii{} luminosities between $10^9$ and $10^{10}$\,\Lsun{}. The typical \Cii{} line width is $\sim385$\,\kms{}, similar to what has been found by previous studies.
\item[{\em ii-}] The \Cii{}/FIR luminosity ratio (and the \Cii{} equivalent width) range over $\sim1$ dex, from very low ratios as in local ULIRGs up to higher ratios typical of normal star--forming galaxies. Despite the relatively low angular resolution of our observations, we find a dependence of the \Cii{}/FIR ratio on the surface brightness of the IR emission, which mimics the results from low--redshift IR--luminous galaxies. In the low--$z$ comparison sample considered here, this ratio also depends on the dust temperature, which varies from $\sim$35\,K to $\sim$50\,K for \Cii{}/FIR ratios comparable to those of the quasar host galaxies in our sample.
\item[{\em iii-}] We infer star formation rates and star formation surface densities $\Sigma_{\rm SFR}$ from both \Cii{} and IR luminosities. None of our quasar host galaxies appears to have $\Sigma_{\rm SFR}$ approaching the Eddington limit, but the modest angular resolution of our data might be biasing our $\Sigma_{\rm SFR}$ low.
\item[{\em iv-}] Four sources are clearly spatially resolved, i.e., they have observed sizes of the \Cii{}--emitting gas that are $\sim1.4\times$ larger than the synthesized beam, and S/N$>$10. Two of them show clear velocity gradients. It is unclear whether these systems are dominated by rotation, or by a more complex dynamical pattern, or if the host galaxies have a \Cii{}--emitting companion at a few kpc separation.  
\item[{\em v-}] A rough estimate of the dynamical mass of the host galaxies in our sample gives values between  $2\times 10^{10}$ and $2\times10^{11}$\,\Msun{}. By assuming Eddington--limited accretion, we estimate a minimum mass of the black holes powering the quasars in our sample, $M_{\rm BH}^{\rm min}=(0.3-3)\times 10^9$\,\Msun. The inferred mass ratio between the black holes and their host galaxies is thus $>15\times$ higher than that of local galaxies \citep[e.g.,][]{kormendy13}, consistent with previous findings based on CO \citep[e.g.,][]{walter04}.
\item[{\em vi-}] By stacking the spectra of all the \Cii{}--detected quasar host galaxies in our sample, we put stringent limits on other lines (H$_2$O, OH+, [N{\sc iii}], [N{\sc iv}], NH$_3$). We do not detect any  deviation from a Gaussian line profile for the \Cii{} line. The strong outflow reported for an individual high--redshift quasar \citep[J1148+5251,][]{cicone15} thus does not appear to be a common feature of $z>6$ quasar hosts.
\item[{\em vii-}] The (rest-frame) UV luminosity of the quasar shows only a weak correlation with the \Cii{} luminosity, and no correlation at all with the \Cii{}/FIR ratio.  
\end{itemize}

The 8 min on--source snapshot observations presented here demonstrate that bright dust and \Cii{} emission are ubiquitous in quasar host galaxies at the highest redshifts currently accessible. Their brightness implies that early chemical enrichment in the hosts was a common phenomenon in the first Gyr of the Universe. Given their enormous flux densities, these sources are unique targets for future ALMA follow-up observations to obtain higher-resolution imaging of the ISM in the host galaxies, and to constrain the physical conditions in the ISM through multi-line observations.

\begin{table*}
\caption{{\rm The main sample of this study.
(1) quasar name. (2) Short name. (3-4) Right ascension and declination (J2000).  
(5) Optical/NIR redshift.
(6) Method for optical/NIR redshift determination.
(7) Redshift reference: 
 1- \citet{banados16};  
 2- Venemans et al.~(in prep);
 3- \citet{carnall15};
 4- \citet{reed17};
 5- \citet{derosa11};
 6- \citet{kurk07};
 7- \citet{venemans15};
 8- \citet{jiang16};    
 9- \citet{matsuoka16};
10- \citet{willott10};
11- \citet{mazzucchelli17};
12- \citet{kurk09};
13- \citet{becker15};
14- \citet{willott09}.
(8) Absolute magnitude at rest-frame 1450 \AA{}.
}} \label{tab_sample}
\begin{center}
\begin{tabular}{cccccccc}
\hline
    Name                     & Short name &   RA         &   Dec        & $z$                  & Method    & Ref. & $M_{1450}$ \\
                             &            & J2000.0	 &   J2000.0    &                      &           &      & [mag]      \\
    (1)                      &    (2)     &    (3)	 &  (4)         &        (5)	       &      (6)  &  (7) &  (8)       \\
    \hline
    PSO~J007.0273+04.9571    &   P007+04   & 00:28:06.560 & +04:57:25.68 & $6.00\pm   0.05  $  &  Template &  1  &     -26.64  \\
    PSO~J009.7355--10.4316   &  P009--10   & 00:38:56.522 & -10:25:53.90 & $5.95\pm   0.05  $  &  Template &  1  &     -26.53  \\
       VIK~J0046--2837       & J0046--2837 & 00:46:23.65  & -28:37:47.34 & $5.99\pm   0.05  $  &  Template &  2  &     -25.48  \\
   ATLAS~J025.6821--33.4627  & J0142--3327 & 01:42:43.73  & -33:27:45.47 & $6.31\pm   0.03  $  &Ly$\alpha$ &  3  &     -27.81  \\
    PSO~J065.4085--26.9543   &  P065--26   & 04:21:38.052 & -26:57:15.60 & $6.14\pm   0.05  $  &  Template &  1  &     -27.25  \\
    PSO~J065.5041--19.4579   &  P065--19   & 04:22:00.994 & -19:27:28.68 & $6.12\pm   0.05  $  &  Template &  1  &     -26.62  \\
       VDESJ0454--4448       & J0454--4448 & 04:54:01.79  & -44:48:31.1  & $6.10\pm   0.01  $  &  Template &  4  &     -26.41  \\
       SDSS~J0842+1218       & J0842+1218  & 08:42:29.429 & +12:18:50.50 & $6.069\pm  0.002 $  &    MgII   &  5  &     -26.91  \\
       SDSS~J1030+0524       & J1030+0524  & 10:30:27.098 & +05:24:55.00 & $6.308\pm  0.001 $  &    MgII   &  6  &     -26.99  \\
    PSO~J159.2257--02.5438   &  P159--02   & 10:36:54.191 & -02:32:37.94 & $6.38\pm   0.05  $  &  Template &  1  &     -26.80  \\
       VIK~J1048--0109       & J1048--0109 & 10:48:19.086 & -01:09:40.29 & $6.661\pm 0.005  $  &    MgII   &  2  &     -26.00  \\
    PSO~J167.6415--13.4960   &  P167--13   & 11:10:33.976 & -13:29:45.60 & $6.508\pm  0.001 $  &    MgII   &  7  &     -25.62  \\
       ULAS~J1148+0702       & J1148+0702  & 11:48:03.286 & +07:02:08.33 & $6.339\pm  0.001 $  &    MgII   &  8  &     -26.49  \\
        VIK~J1152+0055       & J1152+0055  & 11:52:21.269 & +00:55:36.69 & $6.37\pm 0.01    $  &Ly$\alpha$ &  9  &     -25.13  \\
       ULAS~J1207+0630       & J1207+0630  & 12:07:37.440 & +06:30:10.37 & $6.040\pm  0.003 $  &Ly$\alpha$ &  8  &     -26.63  \\
    PSO~J183.1124+05.0926    &   P183+05   & 12:12:26.981 & +05:05:33.49 & $6.4039\pm0.00001$  &     DLA   &  1  &     -27.03  \\
       SDSS~J1306+0356       & J1306+0356  & 13:06:08.258 & +03:56:26.30 & $6.016\pm  0.002 $  &    MgII   &  6  &     -26.81  \\
    PSO~J217.0891--16.0453   &  P217--16   & 14:28:21.394 & -16:02:43.29 & $6.11\pm   0.05  $  &  Template &  1  &     -26.93  \\
      CFHQS~J1509--1749      & J1509--1749 & 15:09:41.778 & -17:49:26.80 & $6.121\pm  0.002 $  &    MgII   & 10  &     -27.14  \\
    PSO~J231.6576--20.8335   &  P231--20   & 15:26:37.841 & -20:50:00.66 & $6.595\pm 0.015  $  &    MgII   & 11  &     -27.20  \\
    PSO~J308.0416--21.2339   &  P308--21   & 20:32:09.996 & -21:14:02.31 & $6.24\pm   0.05  $  &  Template &  1  &     -26.35  \\
      CFHQS~J2100--1715      & J2100--1715 & 21:00:54.616 & -17:15:22.50 & $6.087\pm  0.005 $  &    MgII   & 10  &     -25.55  \\
       VIK~J2211--3206       & J2211--3206 & 22:11:12.391 & -32:06:12.94 & $6.336\pm 0.005  $  &    MgII   &  2  &     -26.71  \\
    PSO~J340.2041--18.6621   &  P340--18   & 22:40:48.997 & -18:39:43.81 & $6.01\pm   0.05  $  &  Template &  1  &     -26.42  \\
       VIK~J2318--3113       & J2318--3113 & 23:18:18.351 & -31:13:46.35 & $6.444\pm 0.005  $  &    MgII   &  2  &     -26.11  \\
       VIK~J2318--3029       & J2318--3029 & 23:18:33.100 & -30:29:33.37 & $6.12\pm 0.05    $  &  Template &  2  &     -26.21  \\
    PSO~J359.1352--06.3831   &  P359--06   & 23:56:32.455 & -06:22:59.26 & $6.15\pm   0.05  $  &  Template &  1  &     -26.79  \\
\hline
\end{tabular}
\end{center}
\end{table*}

\begin{table*}
\caption{{\rm The sample from the literature. 
(1) quasar name. (2-3) Right ascension and declination (J2000).  
(4) Redshift.
(5) Absolute magnitude at 1450 \AA{} rest-frame.
(6) \Cii{} luminosity.
(7) \Cii{} Full Width at Half Maximum.
(8) Total IR luminosity.
(9) Reference: 
   1- \citet{mazzucchelli17}; 
   2- \citet{willott15}; 
   3- \citet{wang16};
   4- \citet{venemans16};
   5- \citet{wang13}; 
   6- \citet{willott13};
   7- \citet{banados15};
   8- \citet{venemans12};
   9- \citet{walter09};
  10- \citet{willott17};
  11- \citet{venemans17c}.
}} \label{tab_lit}
\begin{center}
\begin{tabular}{ccccccccc}
\hline
    Name    &   RA         &   Dec        & $z$      & $M_{1450}$ &    $L_{\rm [CII]}$	  &   FWHM(\Cii)           &	  log $L_{\rm TIR}$         & Ref \\
            &  J2000.0	   &   J2000.0    &          & [mag]      &[$\times10^9$~\Lsun{}] & [\kms{}]               &	     [\Lsun{}]    	    &     \\
    (1)     &    (2)       &    (3)	  &  (4)     &	(5)	  &      (6)		  &    (7)		   &            (8)                 & (9) \\
    \hline
   P006+39   & 00:24:29.772 & +39:13:18.98  & 6.6100   &  -26.9	    &	$0.9   \pm 0.5  $ & $ 280     \pm     140$ & $ 12.05 \pm   0.17 $ &   1 \\
 J0055+0146  & 00:55:02.910 & +01:46:18.30  & 6.0060   &  -24.81    &	$0.827 \pm 0.13 $ & $ 359     \pm     45 $ & $ 11.58 \pm   0.08 $ &   2 \\
 J0100+2802  & 01:00:13.027 & +28:02:25.84  & 6.3258   &  -29.14    &	$3.56  \pm 0.49 $ & $ 300     \pm     77 $ & $ 12.42 \pm   0.09 $ &   3 \\
 J0109--3047 & 01:09:53.131 & --30:47:26.32 & 6.7909   &  -25.64    &	$ 2.4  \pm  0.2 $ & $ 340     \pm     36 $ & $ 12.08 \pm   0.09 $ &   4 \\
 J0129--0035 & 01:29:58.510 & --00:35:39.70 & 5.7787   &  -23.89    &	$ 1.9  \pm  0.3 $ & $ 194     \pm     12 $ & $ 12.64 \pm   0.010$ &   5 \\
 J0210--0456 & 02:10:13.190 & --04:56:20.90 & 6.4323   &  -24.53    &	$0.29  \pm 0.041$ & $ 189     \pm     18 $ & $ 11.38 \pm   0.15 $ &   6 \\
   P036+03   & 02:26:01.876 & +03:02:59.39  & 6.5412   &  -27.33    &	$ 5.8  \pm  0.7 $ & $ 360     \pm     50 $ & $ 12.71 \pm   0.11 $ &   7 \\
 J0305--3150 & 03:05:16.916 & --31:50:55.90 & 6.6145   &  -26.18    &	$ 3.9  \pm  0.2 $ & $ 255     \pm     12 $ & $ 12.83 \pm   0.013$ &   4 \\
 J1044--0125 & 10:44:33.042 & --01:25:02.20 & 5.7847   &  -27.38    &	$ 1.6  \pm  0.4 $ & $ 420     \pm     80 $ & $ 12.72 \pm   0.013$ &   5 \\
 J1120+0641  & 11:20:01.479 & +06:41:24.30  & 7.0842   &  -26.63    &	$ 1.3  \pm  0.2 $ & $ 235     \pm     35 $ & $ 12.15 \pm   0.13 $ &   8 \\
 J1148+5251  & 11:48:16.652 & +52:51:50.44  & 6.4189   &  -27.62    &	$ 4.2  \pm 0.35 $ & $ 287     \pm     28 $ & $ 13.00 \pm   0.06 $ &   9 \\
 J1319+0950  & 13:19:11.302 & +09:50:51.49  & 6.1330   &  -27.05    &	$ 4.4  \pm  0.9 $ & $ 515     \pm     81 $ & $ 12.99 \pm   0.008$ &   5 \\
 J1342+0928  & 13:42:08.100 & +09:28:38.61  & 7.5413   &  -26.76    &   $ 1.6  \pm  0.2 $ & $ 383     \pm     56 $ & $ 12.10 \pm   0.20 $ &  11 \\
 J2054--0005 & 20:54:06.481 & --00:05:14.80 & 6.0391   &  -26.21    &	$ 3.3  \pm  0.5 $ & $ 243     \pm     10 $ & $ 12.73 \pm   0.007$ &   5 \\
   P323+12   & 21:32:33.191 & +12:17:55.26  & 6.5850   &  -27.12    &	$ 1.2  \pm  0.3 $ & $ 250     \pm     30 $ & $ 11.99 \pm   0.16 $ &   1 \\
 VIMOS2911   & 22:19:17.227 & +01:02:48.88  & 6.1492   &  -23.10    &   $2.59  \pm 0.13 $ & $ 264     \pm     15 $ & $ 12.24 \pm   0.03 $ &  10 \\
 J2229+1457  & 22:29:01.649 & +14:57:08.99  & 6.1517   &  -24.78    &	$0.594 \pm 0.077$ & $ 351     \pm     39 $ & $ 11.00 \pm   0.33 $ &   2 \\
   P338+29   & 22:32:55.150 & +29:30:32.23  & 6.6580   &  -26.14    &	$ 2.0  \pm  0.1 $ & $ 740     \pm     310$ & $ 12.31 \pm   0.11 $ &   1 \\
 J2310+1855  & 23:10:38.882 & +18:55:19.70  & 6.0031   &  -27.8	    &	$ 8.7  \pm  1.4 $ & $ 393     \pm     21 $ & $ 13.20 \pm   0.004$ &   5 \\
 J2329--0301 & 23:29:08.275 & --03:01:58.80 & 6.4170   &  -25.25    &	$ 0.39 \pm 0.04 $ & $ 477     \pm     64 $ & $ 10.95_{-0.30}^{+0.17}$ &  10 \\
 J2348--3054 & 23:48:33.334 & --30:54:10.24 & 6.9018   &  -25.8	    &	$ 1.9  \pm  0.3 $ & $ 405     \pm     69 $ & $ 12.63 \pm   0.03 $ &   4 \\
\hline
\end{tabular}
\end{center}
\end{table*}

\begin{table*}[p]
\caption{Observing log, integration time, beam (major $\times$ minor axis, Position Angle), and noise rms (computed as the median of the rms values measured in the SPW0\&1 channels, with 30\,\kms{} binning).} \label{tab_obs}
\begin{center}
\begin{tabular}{cccccccc}
\hline
 Short name &  Date Obs. & Exp.Time & Beam          & Beam PA &	rms (30\,\kms{}) \\ 
            &            & [min]    & [$''$]	    & [deg]   & [mJy\,beam$^{-1}$] \\
    (1)     &   (2)      &   (3)    &  (4)	    &   (5)   &	   (6) \\
    \hline
PJ007+04   & 19 Jun 2016 & 7.56  &     $0.66\times0.46$ &  66.0 &  0.64 \\
PJ009--10  & 19 Jun 2016 & 8.57  &     $0.63\times0.43$ &  84.2 &  0.59 \\
J0046--2837& 07 Jul 2016 & 8.57  &     $0.49\times0.41$ & -86.9 &  0.49 \\
J0142--3327& 14 Apr 2016 & 8.06  &     $0.84\times0.73$ & -49.4 &  0.54 \\
PJ065--26  & 29 Jan 2016 & 8.06  &     $1.10\times0.83$ &  89.1 &  0.59 \\
	   & 31 Mar 2016 & 8.06  &		       &       &       \\
PJ065--19  & 01 Feb 2016 & 8.06  &     $1.07\times0.71$ & -78.4 &  0.47 \\
	   & 09 Apr 2016 & 8.06  &		       &       &       \\
J0454--4448& 29 Jan 2016 & 8.57  &     $1.13\times0.77$ &  89.0 &  0.47 \\
	   & 22 Mar 2016 & 8.57  &		       &       &       \\
J0842+1218 & 31 Jan 2016 & 7.56  &     $1.20\times1.06$ &  77.5 &  0.65 \\
J1030+0524 & 29 Jan 2016 & 8.57  &     $1.16\times0.94$ &  64.0 &  0.58 \\
PJ159--02  & 29 Jan 2016 & 8.57  &     $1.23\times0.94$ &  67.6 &  0.44 \\
J1048--0109& 29 Jan 2016 & 12.10 &     $1.40\times0.97$ &  64.0 &  0.42 \\
PJ167--13  & 29 Jan 2016 & 8.06  &     $1.23\times0.93$ &  76.1 &  0.42 \\
J1148+0702 & 30 Jan 2016 & 8.57  &     $1.25\times1.11$ &  87.0 &  0.55 \\
J1152+0055 & 30 Jan 2016 & 8.06  &     $1.23\times0.99$ & -85.4 &  0.56 \\
J1207+0630 & 31 Jan 2016 & 7.56  &     $1.54\times0.83$ &  57.8 &  0.65 \\
PJ183+05   & 27 Jan 2016 & 8.57  &     $1.19\times1.00$ & -89.7 &  0.50 \\
J1306+0356 & 27 Jan 2016 & 8.57  &     $1.11\times0.91$ &  74.4 &  0.56 \\
PJ217--16  & 27 Jan 2016 & 8.57  &     $1.15\times0.88$ &  78.4 &  0.64 \\
J1509--1749& 31 Jan 2016 & 8.06  &     $1.35\times0.87$ &  65.0 &  0.50 \\
PJ231--20  & 27 Jan 2016 & 7.56  &     $1.24\times0.89$ &  74.9 &  0.58 \\
PJ308--21  & 27 Mar 2016 & 12.60 &     $0.85\times0.65$ &  79.2 &  0.46 \\
J2100--1715& 26 Mar 2016 & 8.06  &     $0.74\times0.63$ & -86.5 &  0.72 \\
J2211--3206& 30 Mar 2016 & 8.06  &     $0.87\times0.70$ &  83.1 &  0.44 \\
PJ340--18  & 09 Apr 2016 & 8.06  &     $0.78\times0.66$ &  84.4 &  0.59 \\
J2318--3113& 02 Apr 2016 & 8.06  &     $0.83\times0.77$ & -87.2 &  1.04 \\
J2318--3029& 12 Apr 2016 & 8.06  &     $0.86\times0.74$ & -53.1 &  0.84 \\
PJ359--06  & 27 Apr 2016 & 8.06  &     $1.08\times0.61$ &  67.4 &  0.86 \\
\hline
\end{tabular}
\end{center}
\end{table*}

\begin{table*}
\caption{{\rm Results from the spectral fit. These measurements are not corrected for extended emission.
(1) Quasar name. 
(2) Line peak frequency.
(3) Inferred \Cii{} redshift.
(4) Line Full Width at Half Maximum.
(5) Integrated line flux.
(6) Continuum flux density at 158$\mu$m (rest frame).
}} \label{tab_spc}
\begin{center}
\begin{tabular}{ccccccccc}
\hline
Short Name  &   $\nu_{\rm obs}$(\Cii)     &   $z_{\rm [CII]}$               & FWHM(\Cii)           & $F_{\rm line}$(\Cii)        & $F_\nu$(cont)               \\
            &   [GHz]                     &                                 & [\kms]               & [Jy\,\kms{}]                & [mJy]                       \\
    (1)     &    (2)                      &    (3)	                    &  (4)                 &	(5)	                 &      (6)                    \\
    \hline
     P007+04   & $271.478_{-0.016}^{+0.015}$ & $6.0008_{-0.0004}^{+0.0004}$ & $ 340_{- 36}^{+ 36}$ & $ 1.58_{-0.07}^{+0.08}$ & $  2.07_{-0.04}^{+0.04}$ \\
    P009-10    & $271.356_{-0.016}^{+0.015}$ & $6.0039_{-0.0004}^{+0.0004}$ & $ 251_{- 34}^{+ 34}$ & $ 2.49_{-0.15}^{+0.15}$ & $  1.89_{-0.07}^{+0.03}$ \\
   J0046--2837 &  ---		             &  ---		            &  ---		   &  ---                    & $  0.18_{-0.04}^{+0.04}$ \\
   J0142--3327 & $259.004_{-0.015}^{+0.014}$ & $6.3379_{-0.0004}^{+0.0004}$ & $ 300_{- 31}^{+ 32}$ & $ 2.62_{-0.06}^{+0.06}$ & $  1.65_{-0.04}^{+0.04}$ \\
    P065--26   & $264.417_{-0.018}^{+0.018}$ & $6.1877_{-0.0005}^{+0.0005}$ & $ 517_{- 43}^{+ 44}$ & $ 2.05_{-0.11}^{+0.10}$ & $  1.23_{-0.05}^{+0.05}$ \\
    P065--19   & $266.753_{-0.023}^{+0.021}$ & $6.1247_{-0.0006}^{+0.0006}$ & $ 345_{- 61}^{+ 67}$ & $ 0.69_{-0.08}^{+0.08}$ & $  0.46_{-0.04}^{+0.05}$ \\
   J0454--4448 & $269.272_{-0.022}^{+0.022}$ & $6.0581_{-0.0006}^{+0.0006}$ & $ 426_{- 55}^{+ 57}$ & $ 0.85_{-0.07}^{+0.08}$ & $  0.71_{-0.05}^{+0.05}$ \\
   J0842+1218  & $268.580_{-0.019}^{+0.019}$ & $6.0763_{-0.0005}^{+0.0005}$ & $ 396_{- 44}^{+ 45}$ & $ 1.44_{-0.10}^{+0.11}$ & $  0.65_{-0.05}^{+0.06}$ \\
   J1030+0524  &  ---		             &  ---		            &  ---		   &  ---                    & $  0.03_{-0.05}^{+0.06}$ \\ 
    P159--02   & $257.496_{-0.017}^{+0.017}$ & $6.3809_{-0.0005}^{+0.0005}$ & $ 373_{- 39}^{+ 40}$ & $ 1.15_{-0.07}^{+0.07}$ & $  0.65_{-0.03}^{+0.03}$ \\
   J1048--0109 & $247.598_{-0.015}^{+0.014}$ & $6.6759_{-0.0004}^{+0.0005}$ & $ 330_{- 33}^{+ 32}$ & $ 2.52_{-0.06}^{+0.07}$ & $  2.84_{-0.04}^{+0.03}$ \\
    P167--13   & $252.907_{-0.016}^{+0.014}$ & $6.5148_{-0.0004}^{+0.0005}$ & $ 437_{- 34}^{+ 34}$ & $ 2.53_{-0.07}^{+0.07}$ & $  0.87_{-0.05}^{+0.02}$ \\
   J1148+0702  &  ---		             &  ---		            &  ---		   &  ---                    & $  0.41_{-0.05}^{+0.05}$ \\ %
   J1152+0055  & $258.076_{-0.017}^{+0.016}$ & $6.3643_{-0.0004}^{+0.0005}$ & $ 167_{- 44}^{+ 45}$ & $ 0.54_{-0.06}^{+0.07}$ & $  0.22_{-0.05}^{+0.04}$ \\
   J1207+0630  & $270.094_{-0.032}^{+0.034}$ & $6.0366_{-0.0009}^{+0.0008}$ & $ 489_{- 72}^{+ 78}$ & $ 0.92_{-0.12}^{+0.11}$ & $  0.50_{-0.06}^{+0.06}$ \\
     P183+05   & $255.497_{-0.015}^{+0.013}$ & $6.4386_{-0.0004}^{+0.0004}$ & $ 374_{- 30}^{+ 30}$ & $ 5.84_{-0.08}^{+0.08}$ & $  4.47_{-0.02}^{+0.02}$ \\
   J1306+0356  & $270.207_{-0.015}^{+0.014}$ & $6.0337_{-0.0004}^{+0.0004}$ & $ 246_{- 31}^{+ 31}$ & $ 1.63_{-0.09}^{+0.09}$ & $  0.94_{-0.05}^{+0.07}$ \\
    P217--16   & $265.817_{-0.040}^{+0.038}$ & $6.1498_{-0.0010}^{+0.0011}$ & $ 491_{- 75}^{+ 74}$ & $ 0.70_{-0.11}^{+0.11}$ & $  0.37_{-0.06}^{+0.06}$ \\
   J1509--1749 & $266.838_{-0.024}^{+0.024}$ & $6.1225_{-0.0006}^{+0.0007}$ & $ 631_{- 68}^{+ 72}$ & $ 1.50_{-0.12}^{+0.11}$ & $  1.72_{-0.05}^{+0.05}$ \\
    P231--20   & $250.520_{-0.016}^{+0.015}$ & $6.5864_{-0.0005}^{+0.0005}$ & $ 404_{- 37}^{+ 39}$ & $ 2.65_{-0.12}^{+0.11}$ & $  3.36_{-0.04}^{+0.05}$ \\
    P308--21   & $262.720_{-0.019}^{+0.018}$ & $6.2341_{-0.0005}^{+0.0005}$ & $ 570_{- 43}^{+ 45}$ & $ 1.79_{-0.08}^{+0.10}$ & $  0.64_{-0.05}^{+0.03}$ \\
   J2100--1715 & $268.393_{-0.020}^{+0.018}$ & $6.0812_{-0.0005}^{+0.0005}$ & $ 382_{- 47}^{+ 51}$ & $ 1.52_{-0.14}^{+0.14}$ & $  0.52_{-0.06}^{+0.06}$ \\
   J2211--3206 & $258.952_{-0.037}^{+0.036}$ & $6.3394_{-0.0010}^{+0.0010}$ & $ 529_{-100}^{+118}$ & $ 0.57_{-0.11}^{+0.11}$ & $  0.57_{-0.04}^{+0.05}$ \\
    P340--18   &  ---		             &  ---		            &  ---		   &  ---                    & $  0.13_{-0.05}^{+0.05}$ \\ 
   J2318--3113 & $255.330_{-0.019}^{+0.018}$ & $6.4435_{-0.0005}^{+0.0005}$ & $ 234_{- 49}^{+ 49}$ & $ 1.11_{-0.14}^{+0.13}$ & $  0.57_{-0.08}^{+0.09}$ \\
   J2318--3029 & $265.968_{-0.016}^{+0.015}$ & $6.1458_{-0.0004}^{+0.0004}$ & $ 320_{- 34}^{+ 33}$ & $ 2.34_{-0.11}^{+0.12}$ & $  2.71_{-0.08}^{+0.07}$ \\
    P359--06   & $264.988_{-0.017}^{+0.015}$ & $6.1722_{-0.0004}^{+0.0004}$ & $ 330_{- 37}^{+ 39}$ & $ 2.47_{-0.16}^{+0.13}$ & $  0.87_{-0.07}^{+0.09}$ \\
\hline
\end{tabular}
\end{center}
\end{table*}

\begin{table*}
\caption{Results from the 2D gaussian fit of the \Cii{} continuum--subtracted line maps. (1) quasar name. (2) S/N of the line detection. (3) Observed (= beam-convolved) size of the \Cii{}--emitting region from 2D gaussian fit of the continuum--subtracted line maps. (4) Beam-deconvolved size of the \Cii{}--emitting region. (5) Radius of the \Cii{} emission. (6) Measured \Cii{} flux from the 2D gaussian fit. (7) \Cii{} flux from the 2D gaussian fit, corrected to account for the flux loss due to the line wings not covered in the line maps. (8) \Cii{} luminosity. (9) FIR luminosity. (10) FIR surface luminosity. (11) \Cii{} Equivalent Width, in \kms{}.
} \label{tab_size}
\begin{center}
\begin{tabular}{ccccccccccc}
\hline
 Short name & S/N & \Cii{} size        & \Cii{} dec. size & $R_{\rm [CII]}$ & $F_{\rm [CII]}$(2D) & $F_{\rm [CII]}$(corr) & log $L_{\rm [CII]}$ & log $L_{\rm FIR}$ & log $\Sigma_{\rm FIR}$ & log EW$_{\rm [CII]}$\\
            &(\Cii)& [$''$]	       & [$''$]		     & [kpc] & [Jy\,\kms{}] & [Jy\,\kms{}] & [\Lsun] & [\Lsun] & [\Lsun{}\,kpc$^{-2}$] & [\kms]\\
    (1)     &(2)  &  (3)	       & (4)		     & (5)& (6)             & (7)          & (8)& (9)   & (10)   & (11) \\
    \hline
PJ007+04    &  21 & $0.84\times0.60$ & $0.54\times0.34$ &1.6& $1.87\pm0.26$ & $2.24\pm0.31$ & 9.25 & 12.91 & 11.80 & $2.76_{0.07}^{0.06}   $   \\
PJ009--10   &  17 & $0.99\times0.72$ & $0.82\times0.47$ &2.4& $5.26\pm0.39$ & $6.31\pm0.47$ & 9.70 & 12.97 & 11.55 & $3.15_{0.05}^{0.04}   $   \\
J0046--2837 & --- &		 --- &      ---	        &---& ---	    & ---	    & ---  & 11.46 &  ---  &   ---                     \\
J0142--3327 &  41 & $1.28\times0.90$ & $0.98\times0.48$ &2.7& $3.57\pm0.20$ & $4.29\pm0.24$ & 9.56 & 12.78 & 11.33 & $3.23_{0.04}^{0.04}   $   \\
PJ065--26   &  18 & $1.41\times1.02$ & $0.90\times0.56$ &2.5& $2.36\pm0.25$ & $2.83\pm0.30$ & 9.37 & 12.57 & 11.15 & $3.24_{0.06}^{0.05}   $   \\
PJ065--19   &   9 & $1.69\times0.76$ & $1.32\times0.26$ &3.7& $0.97\pm0.11$ & $1.16\pm0.13$ & 8.97 & 12.02 & 10.32 & $3.38_{0.11}^{0.09}   $   \\
J0454--4448 &  12 & $1.38\times1.04$ & $0.92\times0.50$ &2.6& $0.76\pm0.16$ & $0.91\pm0.19$ & 8.86 & 12.33 & 11.06 & $2.96_{0.12}^{0.10}   $   \\
J0842+1218  &  15 & $1.23\times1.21$ & $0.61\times0.15$ &1.7& $1.17\pm0.11$ & $1.41\pm0.14$ & 9.05 & 12.20 & 10.25 & $3.28_{0.17}^{0.12}   $   \\
J1030+05    &   4 &		 --- &      ---	        &---& ---	    & ---	    & ---  & 11.53 &  ---  &   ---                     \\
PJ159--02   &  17 & $1.62\times1.20$ & $1.12\times0.64$ &3.1& $1.09\pm0.12$ & $1.31\pm0.14$ & 9.05 & 12.21 & 11.06 & $3.29_{0.08}^{0.07}   $   \\
J1048--0109 &  43 & $1.49\times1.13$ & $0.60\times0.49$ &1.6& $2.18\pm0.08$ & $2.62\pm0.09$ & 9.38 & 12.92 & 11.82 & $2.93_{0.02}^{0.02}   $   \\
PJ167--13   &  35 & $1.76\times1.32$ & $1.37\times0.78$ &3.7& $3.23\pm0.18$ & $3.88\pm0.22$ & 9.54 & 12.43 & 10.81 & $3.56_{0.05}^{0.04}   $   \\
J1148+0702  &   3 &		 --- &      ---	        &---& ---	    & ---	    & ---  & 12.19 &  ---  &   ---                     \\
J1152+0055  &   9 & $1.73\times1.46$ & $1.30\times0.98$ &3.6& $0.62\pm0.12$ & $0.75\pm0.14$ & 8.81 & 11.50 &  9.74 & $3.78_{0.22}^{0.14}   $   \\
J1207+0630  &   8 & $1.96\times1.24$ & $1.23\times0.90$ &3.5& $1.40\pm0.33$ & $1.68\pm0.40$ & 9.13 & 11.99 & 10.29 & $3.56_{0.16}^{0.12}   $   \\
PJ183+05    &  70 & $1.35\times1.21$ & $0.74\times0.57$ &2.0& $6.52\pm0.17$ & $7.83\pm0.21$ & 9.83 & 13.16 & 11.90 & $3.13_{0.017}^{0.016} $   \\
J1306+0356  &  17 & $1.77\times1.21$ & $1.41\times0.73$ &4.0& $2.55\pm0.20$ & $3.06\pm0.24$ & 9.38 & 12.50 & 10.76 & $3.32_{0.08}^{0.07}   $   \\
PJ217--16   &   6 & $1.42\times1.07$ & $1.02\times0.16$ &2.9& $1.03\pm0.26$ & $1.23\pm0.31$ & 9.00 & 11.97 &  9.28 & $3.47_{0.18}^{0.13}   $   \\
J1509--1749 &  13 & $1.73\times1.34$ & $1.09\times1.01$ &3.1& $2.42\pm0.37$ & $2.91\pm0.44$ & 9.37 & 12.59 & 10.89 & $3.22_{0.08}^{0.07}   $   \\
PJ231--20   &  23 & $1.42\times1.04$ & $0.80\times0.37$ &2.2& $1.92\pm0.08$ & $2.31\pm0.09$ & 9.32 & 13.04 & 12.51 & $2.74_{0.04}^{0.04}   $   \\
PJ308--21   &  23 & $1.12\times1.01$ & $0.81\times0.69$ &2.3& $1.89\pm0.19$ & $2.26\pm0.23$ & 9.27 & 12.27 & 10.48 & $3.44_{0.13}^{0.10}   $   \\
J2100--1715 &  11 & $1.30\times0.81$ & $1.08\times0.48$ &3.1& $1.80\pm0.23$ & $2.16\pm0.28$ & 9.24 & 11.77 & 10.06 & $3.90_{0.15}^{0.11}   $   \\
J2211--3206 &   5 & $1.64\times1.15$ & $1.47\times0.78$ &4.1& $0.93\pm0.27$ & $1.12\pm0.33$ & 8.98 & 12.24 & 10.56 & $3.18_{0.18}^{0.13}   $   \\
PJ340--18   &   4 &		   --- &      ---	    &---& ---	    & ---	    & ---  & 11.60 &  ---  &   ---                     \\
J2318--3113 &   8 & $1.56\times1.45$ & $1.35\times1.20$ &3.7& $1.70\pm0.47$ & $2.04\pm0.57$ & 9.25 & 12.46 & 11.61 & $2.61_{0.04}^{0.03}   $   \\
J2318--3029 &  21 & $0.93\times0.81$ & $0.44\times0.23$ &1.2& $1.83\pm0.11$ & $2.20\pm0.13$ & 9.25 & 12.92 & 10.29 & $3.69_{0.18}^{0.13}   $   \\
PJ359--06   &  15 & $1.15\times1.06$ & $0.91\times0.29$ &2.6& $3.11\pm0.39$ & $3.73\pm0.47$ & 9.49 & 12.36 & 10.89 & $3.56_{0.12}^{0.09}   $   \\
\hline
\end{tabular}
\end{center}
\end{table*}

\acknowledgements

We are grateful to the referee, prof.~Takeuchi, for his helpful comments. We thank Bade Uzgil for insightful discussions. Support for RD was provided by the DFG priority program 1573 `The physics of the interstellar medium'. FW, BPV, EPF acknowledge support through ERC grant Cosmic\_Dawn. DR acknowledges support from the National Science Foundation under grant number AST-1614213 to Cornell University.
\facility{ALMA} data: 2015.1.01115.S. ALMA is a partnership of ESO (representing its member states), NSF (USA) and NINS (Japan), together with NRC (Canada), NSC and ASIAA (Taiwan), and KASI (Republic of Korea), in cooperation with the Republic of Chile. The Joint ALMA Observatory is operated by ESO, AUI/NRAO and NAOJ.

\label{lastpage}

\end{document}